\documentclass{aa}

\usepackage{graphicx}
\usepackage{txfonts}
\usepackage{natbib}
\bibpunct{(}{)}{;}{a}{}{,}
\usepackage{color}

\begin{document} 

\title{
  Classification of magnetized star--planet interactions: \\
  bow shocks, tails, and inspiraling flows
}

\author{
  Titos Matsakos\inst{1,2} \and
  Ana Uribe\inst{1} \and
  Arieh K\"onigl\inst{1}
}

\authorrunning{T. Matsakos et al.}
\titlerunning{Star--planet interactions}

\institute{
  Department of Astronomy \& Astrophysics, The University of Chicago, Chicago,
    IL 60637, USA \\
  \email{titos.matsakos@uchicago.edu} \and
  LERMA, Observatoire de Paris, Universit\'e Pierre et Marie Curie,
    Ecole Normale Sup\'erieure, Universit\'e Cergy-Pontoise, CNRS, France
}

\date{Received ??; accepted ??}

\abstract
{
Close-in exoplanets interact with their host stars gravitationally as well as
via their magnetized plasma outflows.
The rich dynamics that arises may result in distinct observable features.
}{
Our objective is to study and classify the morphology of the different types of
interaction that can take place between a giant close-in planet (a Hot Jupiter)
and its host star, based on the physical parameters that characterize the
system.
}{
We perform 3D magnetohydrodynamic numerical simulations to model the
star--planet interaction, incorporating a star, a Hot Jupiter, and realistic
stellar and planetary outflows.
We explore a wide range of parameters and analyze the flow structures and
magnetic topologies that develop.
}{
Our study suggests the classification of star--planet interactions into four
general types, based on the relative magnitudes of three characteristic length
scales that quantify the effects of the planetary magnetic field, the planetary
outflow, and the stellar gravitational field in the interaction region.
We describe the dynamics of these interactions and the flow structures that they
give rise to, which include bow shocks, cometary-type tails, and inspiraling
accretion streams.
We point out the distinguishing features of each of the classified cases and 
discuss some of their observationally relevant properties.
}{
The magnetized interactions of star--planet systems can be categorized, and
their general morphologies predicted, based on a set of basic stellar,
planetary, and orbital parameters.
}


\maketitle

\section{Introduction}

The exoplanet class ``Hot Jupiters'' consists of massive gaseous planets (of 
mass $\sim$$1$--$10\,M_\mathrm{J}$, where the subscript ``J'' denotes the planet
Jupiter) that orbit their host stars on close-in trajectories (semi-major axis
of $\lesssim$$0.1\,\mathrm{AU}$, corresponding to an orbital period of a few
days).
Among other exoplanets, the size and location of Hot Jupiters favors both their
discovery and the collection of detailed data.
The two leading and complementary detection techniques, radial velocity and
transit observations, have resulted in a wealth of information regarding their
orbital characteristics, such as obliquity and eccentricity
\citep[e.g.,][]{FabryckyWinn09}, as well as their physical properties, such as
surface temperature and atmospheric composition \citep[e.g.,][]{Bean+13}.
The study of the phenomenology of Hot Jupiters also provides valuable input to
research on the physical conditions and processes in smaller, harder to probe,
Neptune- and Earth-like bodies (see, e.g., \citealt{Seager11} for a recent
reference on exoplanets).

Hot Jupiter atmospheres are believed to be heated by photoionizing
extreme-ultraviolet (EUV) stellar irradiation (e.g., \citealt{Burrows+00}, but
see also \citealt{OwenJackson12} as well as \citealt{Buzasi13} and
\citealt{Lanza13}, who proposed, respectively, X-rays for very young stars and
magnetic reconnection for very close-in planets).
The energy input could be large enough to induce an evaporative mass loss at a
rate of $\sim 10^{10}$--$10^{12}\,\mathrm{g\,s}^{-1}$.
Such outflows have been inferred in a couple of systems (HD 189733b and HD
209458b) in which detailed observations have been carried out
\citep[e.g.,][although see \citealt{Ben-Jaffel07, Ben-Jaffel08} for a different
interpretation of the data]{Vidal-Madjar+04, Ehrenreich+08,
LecavelierDesEtangs+10, LecavelierDesEtangs+12, Bourrier+13}.
The theory of evaporative planetary outflows has been developed in a number of
papers, taking into account the thermal and ionization structure of the
irradiated planetary atmosphere as well as tidal and magnetic effects
\citep[e.g.,][and references therein]{LecavelierDesEtangs+04, Yelle04, Yelle06,
Tian+05, GarciaMunoz07, MurrayClay+09, Adams11, Trammell+11, Trammell+14,
Koskinen+13, OwenAdams14}.

The host stars of close-in planets typically drive magnetized stellar winds that
are accelerated to speeds of a few hundred $\mathrm{km\,s}^{-1}$.
The properties of these winds change as the star evolves, from strong and often
collimated outflows in young stars \citep[e.g.,][]{MattPudritz08,Matt+12a} to
weak and more isotropic winds during the main-sequence phase \citep[e.g.,][and
references therein]{Matt+12b, CohenDrake14}.
The magnetized plasma that is driven out in these winds constitutes the medium
into which the evaporative outflows from close-in expoplanets expand.

A close-in planet and its host stars could thus interact through their
respective outflows and/or magnetic fields.
In either case, the interaction could have a potentially observable signature.
For example, it has been suggested that there might be a detectable bow shock
formed by the supersonic stellar wind in front of the planet (e.g.,
\citealt{Vidotto+10, Vidotto+11a, Vidotto+11b, Vidotto+14, Llama+11, Llama+13,
Ben-JaffelBallester13}), or an observable cometary-type tail associated with the
swept-up planetary wind that trails the planet in its orbit \citep[e.g.,][]
{Schneider+98, Mura+11, Kulow+14}.
According to these proposals, a bow shock could potentially contribute to the
absorption of stellar radiation before an ingress of a transiting Hot Jupiter,
whereas a tail might contribute after an egress.
It was furthermore proposed that the interaction with the stellar magnetosphere
might lead to phased hot spots on the stellar surface or to flaring activity
\citep{Shkolnik+03, Shkolnik+04, Shkolnik+05, Shkolnik+08, Walker+08, Lanza08,
Lanza09, Cohen+09, Pillitteri+10, Pillitteri+11, Pillitteri+14}, and that the
planetary magnetic field could be affected in both its topology and its strength
\citep[e.g.,][]{Laine+08, Cohen+11b, Cohen+14, Strugarek+14}.
It was also suggested \citep{Cohen+10} that the interaction with a magnetized
Hot Jupiter could lead to a reduction in the mass loss rate, and even more
strongly in the angular momentum loss rate, from the host star.

In this paper we perform 3D magnetohydrodynamic (MHD) simulations that aim to
address the following questions:
\begin{enumerate}
\item
How can one classify the different types of star--planet interactions that arise
when a Hot Jupiter orbits its host star?
\item
What are the dynamical features that develop in each case, and what are their
expected observational signatures?
\item
What is the impact of this interaction on the planetary and stellar outflows?
\end{enumerate}
We classify such interactions by studying the evolution of a grid of numerical
models.
In particular, we investigate whether it is the planetary magnetosphere or the
planetary outflow that intercepts the stellar plasma, whether the stellar
outflow terminates in a bow shock, the circumstances for forming a stable tail
along the planet's orbit, and the conditions under which accretion streams of
planetary material reach the stellar surface.
We carry out a parametric study and present the criteria that can be used to
classify a system's morphology and dynamics based on its physical parameters.
We also consider some general properties of the interaction regions that bear on
their potential detectability.

Previous work by \citet{Cohen+09, Cohen+11a} has identified and described some
of the physical processes that may take place in a system consisting of a
magnetized star and a magnetized Hot Jupiter.
These studies were, however, focused on the interpretation of two specific
systems.
In addition, the planetary outflows were not explicitly included; they developed
based on the boundary conditions applied on the surface of the planet.
In contrast, we perform a systematic study to classify the types of star--planet
interactions, incorporating consistent planetary winds that are based on
detailed models.

The paper is structured as follows.
Sect.~\ref{sec:models} explains the methodology and presents the numerical
models.
Sect.~\ref{sec:results} reports the results of the simulations and describes the
dynamics.
Sect.~\ref{sec:classify} classifies the types of interaction and considers some
of their potential observational manifestations.
Finally, Sect.~\ref{sec:summary} summarizes the conclusions of this work.

\section{Numerical modeling}
  \label{sec:models}

As discussed in \citet[][see also \citealt{Tian+05, GarciaMunoz07}]
{MurrayClay+09}, evaporative Hot Jupiter outflows need to be modeled as a fluid
(using a hydrodynamic or an MHD formulation) rather than as a collection of
individual particles that undergo Jeans escape, because the plasma typically
remains collisional beyond the flow's critical point.
\citet{MurrayClay+09} considered an unmagnetized outflow from a Hot Jupiter
exposed to a UV flux near the low end of the range of values that we model (see
Sect.~\ref{subsec:models}), and evaluated the collisional mean free path
$\ell_\mathrm{pp} = 1/(n_{\rm H}\sigma_\mathrm{pp})$ to proton--proton
collisions in a hydrogen plasma of density $n_{\rm H}$ and temperature
$T\approx10^4\,$K (for which the relevant cross section is
$\sigma_{\rm pp} \approx  10^{-13} [T/10^4\, {\rm K}]^{-2}\,{\rm cm^2}$).
They inferred a value ($\simeq 10^7\,$cm) that is $\sim$$10^4$ smaller than the
density scale height at the sonic point (which lies at a distance of $\sim$$5$
planetary radii from the center of the planet).
The large value of this ratio indicates that the collisional approximation
should be applicable to this problem over a wide range of circumstances.
Similarly, our simple stellar-wind model (Sect.~\ref{sec:outflows}) is also
adequately modeled within the hydrodynamic framework \citep[e.g.,][]{Parker60}.
The fact that both the planetary and the stellar outflows are likely magnetized
sharpens these conclusions: for typical parameters, the proton Larmor radius
$\ell_\mathrm{L} = c_\mathrm{s}m_\mathrm{p}c/(eB)$ (where $c_{\rm s}$ is the
speed of sound, $m_{\rm p}$ is the proton mass, $c$ is the speed of light, $e$
is the electron charge, and $B$ is the magnetic field amplitude) is
$\sim 10^2\,(c_{\rm s}/10\,\rm{km\,s}^{-1})(B/1\,{\rm G})^{-1}\,$cm --- much
smaller than $\ell_\mathrm{pp}$.
Given that the region where the planetary and stellar outflows interact is
located at a distance from the source of each of these winds that is comparable
to the distance of its respective critical surface, an MHD treatment of this
interaction should be amply justified.

A numerical simulation that encompasses both the star and the planet needs to
adequately resolve the sizes of both objects.
For the case of a Hot Jupiter orbiting a solar-type host, the characteristic
length scales of the planet and the star differ by roughly an order of
magnitude.
Since 3D numerical simulations are computationally demanding, the best strategy
to model the star--planet interaction is to adopt multiple levels of refinement.
One approach is to configure an adaptive mesh that follows the orbit of the
planet and provides high resolution for its interior and environment.
Another is to set up a static, multiply refined grid and combine it properly
with a corotating frame of reference, so that the location of the planet stays
fixed in the highly resolved region.
Here we adopt the latter approach, which also simplifies the treatment of the
interior of the planet.
In general, a static, multiply refined grid is also appropriate for inclined
orbits, as long as they are circular.

In the following, we denote the star with ``$*$'' and the planet with
``$\circ$''.
The simulations are performed in Cartesian coordinates $(x,\,y,\,z)$, but for
convenience we also use the notation of spherical coordinates
$(R,\,\theta,\,\phi)$.
The planet is assumed to orbit in the $x$-$y$ plane, or, equivalently, at
$\theta = \pi/2$.
The center of the star is placed at the origin,
$(x_*,\,y_*,\,z_*) = (0,\,0,\,0)$, whereas the center of the planet at
$(x_\circ,\,y_\circ,\,z_\circ) = (x_\circ,\,0,\,0)$.
For the initialization we use an extra set of coordinate systems denoted with
primes, $(x',\,y',\,z')$ and $(R',\,\theta',\,\phi')$, with their origins placed
at $(x_\circ,\,y_\circ,\,z_\circ)$ (i.e., centered on the planet).
At each spatial point $(x,\,y,\,z)$ we first calculate
$(x',\,y',\,z') = (x - x_\circ,\,y,\,z)$ and then perform a coordinate
transformation to obtain $(R',\,\theta',\,\phi')$.
In order to keep the center of the planet fixed, we adopt a corotating frame
with $\vec\Omega_\mathrm{fr} = \vec\Omega_\mathrm{orb}$, where
$\vec\Omega_\mathrm{fr} = \Omega_\mathrm{fr}\hat z$ is the angular velocity of
the frame and $\Omega_\mathrm{orb} = (GM_*/x_\circ^3)^{1/2}$ is the Keplerian
angular speed of the planet (with $M_*$ being the mass of the star and $G$ the
gravitational constant).
Since $M_* \gg M_\circ$ (where $M_\circ$ is the mass of the planet), we assume
that $M_* + M_\circ \simeq M_*$ and hence that the origin of the coordinate
system is effectively located at the center of mass.

\subsection{The MHD equations}

The ideal MHD equations in a frame that corotates with the planet are:
\begin{equation}
  \frac{\partial\rho}{\partial t} + \nabla \cdot (\rho \vec v) = 0\,,
  \label{eq:mass}
\end{equation}
\begin{equation}
  \frac{\partial\vec v}{\partial t} + (\vec v \cdot \nabla)\vec v
    + \frac{1}{4\pi\rho}\vec B \times (\nabla \times \vec B)
    + \frac{1}{\rho}\nabla P = \vec g + \vec F_\mathrm{in}\,,
  \label{eq:momentum}
\end{equation}
\begin{equation}
  \frac{\partial P}{\partial t} + \vec v \cdot \nabla P
    + \gamma P \nabla \cdot \vec v = 0\,,
  \label{eq:energy}
\end{equation}
\begin{equation}
  \frac{\partial\vec B}{\partial t} + \nabla \times (\vec B \times \vec v)
    = 0\,.
\end{equation}
The variables $\rho$, $P$, $\vec v$, and $\vec B$ denote the density, pressure,
velocity, and magnetic field, respectively, and $\gamma$ is the polytropic
index.
The magnetic field is required to obey the solenoidal condition
$\nabla\cdot\vec B = 0$.
Since we consider the nonrelativistic regime, $\rho$, $P$, and $\vec B$ have the
same values in the corotating frame and in the lab (star's center of mass)
frame.
The velocity in the lab frame is obtained from the expression
$\vec v_\mathrm{lab} = \vec v + \sin\theta R\Omega_\mathrm{fr}\hat\phi$.
The vector $\vec g = \vec g_* + \vec g_\circ$ is the gravitational acceleration,
and the inertial force that appears in the corotating frame,
$\vec F_\mathrm{in} = \vec F_\mathrm{centr} + \vec F_\mathrm{cor}$, has
centrifugal
\begin{equation}
  \vec F_\mathrm{centr}
    = -\big[\vec\Omega_\mathrm{fr}\times
    (\vec\Omega_\mathrm{fr}\times \vec R)\big]
    = \Omega_\mathrm{fr}^2(x\hat x + y\hat y)
\end{equation}
and Coriolis
\begin{equation}
  \vec F_\mathrm{cor} = -2(\vec\Omega_\mathrm{fr} \times \vec v)
    = 2\Omega_\mathrm{fr}(v_y\hat x - v_x\hat y)
\end{equation}
components.

The temperature at the base of an EUV-induced Hot Jupiter outflow is found to be
$\sim 10^4\,\mathrm{K}$ \citep[e.g.,][]{MurrayClay+09}: this value reflects the
balance between photoionization heating and collisionally excited radiative
cooling \citep[e.g.,][]{Spitzer78}.
Furthermore, in cases where the stellar UV flux is sufficiently high, radiative
cooling continues to roughly balance the photoionization heating as the wind
moves away from the surface, with the result that the temperature changes little
along the flow (e.g., \citealt{MurrayClay+09}).
The isothermal approximation was also found to apply to the inner regions of the
solar wind \citep[e.g.,][]{Cranmer+07}, and has thus been adopted in previous
simulations of both stellar \citep[e.g.,][]{MattPudritz08} and planetary
\citep[e.g.,][]{Trammell+14} outflows.
Based on these considerations, we model both the stellar and the planetary
outflows in our simulations as being nearly isothermal throughout,
characterizing them by a polytropic index $\gamma = 1.05$ as in the fiducial
model of \citet{MattPudritz08}.
This approach enables us to simulate the dynamics of these winds without having
to deal with the intricacies of a detailed calculation of the thermal structure
of the outflow.
As we show in Sect.~\ref{sec:outflows}, the adopted framework can be made to
yield density and velocity profiles for the planetary wind that {are compatible
with the results obtained by employing the more elaborate (albeit nonmagnetic)
1D model of \citet{MurrayClay+09}, which incorporates photoionization heating
and radiative as well as adiabatic cooling.
\cite{Koskinen+13} --- who adopted a more detailed modeling approach that
included, among other factors, a consideration of the full spectrum of ionizing
photons (rather than just a single representative frequency) --- found
quantitative differences from the \citet{MurrayClay+09} findings (such as a
slower increase of the ionization fraction with distance from the planet's
surface; see also \citealt{Trammell+11}), but their results remain qualitatively
similar to those of \citet{MurrayClay+09}.
We therefore consider the adopted approximation to be justified for the present
study, especially in view of the fact that the classification criteria presented
in Sect.~\ref{sec:classify} do not depend on the details of the wind models.

We further assume, for simplicity, that the plasma consists of pure atomic
hydrogen, and that it is fully ionized, so that the temperature is given by
$T = P m_\mathrm{p}/2\rho k_\mathrm{B}$ (where the factor $1/2$ represents the
mean molecular weight, and $k_\mathrm{B}$ is the Boltzmann constant).
The strong ionization and near-isothermality assumptions for the planetary wind
are consistent with the ``high UV flux'' case (corresponding to a young host
star; e.g., \citealt{Ribas+05}) presented in \citet{MurrayClay+09}.
By contrast, in the ``low UV flux'' case considered by these authors
(corresponding to a solar-analog host), the wind temperature drops with distance
from the stellar surface on account of adiabatic cooling.
To mimic this case, we also consider models with a characteristic temperature
that is lower than $10^4\,\mathrm{K}$. 

At locations where the magnetized stellar wind interacts with the planetary
outflow or magnetosphere, field-line reconnection may potentially take place.
This process can be realized in the simulations --- even though resistivity is
not explicitly included in the formulation --- because numerical diffusion
allows the magnetic field to reconnect in the presence of strong current sheets.
We note, however, that the model setup adopted in this paper represents the star
and the planet as aligned magnetic dipoles, a configuration that does not give
rise to X-point structures and is therefore not prone to reconnection.

\subsection{Stellar and planetary parameters}
\label{sec:parameters}

\begin{table*}
  \caption{
    Range of parameters considered in the simulations.
  \label{tab:parameters}}
  \centering
  \begin{tabular}{lcccc}
    \hline
    \hline
    Parameter
      & Symbol
      & Stellar value          & Planetary value                      & Units                                   \\
    \hline
    Radius
      & $R_*$, $R_\circ$
      & $1\,R_{\sun}$          & $1$--$1.5\,R_\mathrm{J}$             & $R_\mathrm{J} \simeq 10^{-1}\,R_{\sun}$ \\
    Mass
      & $M_*$, $M_\circ$
      & $1\,M_{\sun}$          & $0.5$--$1\,M_\mathrm{J}$             & $M_\mathrm{J} \simeq 10^{-3}\,M_{\sun}$ \\
    Temperature at the base of the outflow
      & $T_*$, $T_\circ$
      & $10^6$                 & $6\times10^3$--$10^4$                & K                                       \\
    Density at the base of the outflow
      & $\rho_*$, $\rho_\circ$
      & $5\times10^{-15}$      & $7\times10^{-17}$--$3\times10^{-13}$ & g\,cm$^{-3}$                            \\ 
    Equatorial surface magnetic field
      & $B_*$, $B_\circ$
      & $2$                    & $0.1$--$1$                           & G                                       \\
    Escape speed at the base of the outflow
      & $v_{\mathrm{esc}\,*}$, $v_{\mathrm{esc}\,\circ}$
      & $620$                  & $35$--$60$                           & km\,s$^{-1}$                            \\
    Sound speed at the base of the outflow
      & $c_{\mathrm{s}\,*}$, $c_{\mathrm{s}\,\circ}$
      & $130$                  & $9$--$13$                            & km\,s$^{-1}$                            \\
    Escape velocity parameter $(1/2)(v_\mathrm{esc}/c_\mathrm{s})^2$
      & $\lambda_*$, $\lambda_\circ$
      & $11.5$                 & $3.8$--$23$                          & --                                      \\
    Plasma $\beta$ $[P/(B^2/8\pi)]$ at the base of the outflow
      & $\beta_*$, $\beta_\circ$
      & $5$                    & $0.002$--$400$                       & --                                      \\
    Rotation period
      & $\mathcal{P}_*$, $\mathcal{P}_\circ$
      & $12$                   & $1.2$--$3.7$                         & days                                    \\
    Rotation speed parameter $(\Omega R/c_\mathrm{s})^2$
      & $\epsilon_*$, $\epsilon_\circ$
      & $0.1$                  & $0.03$--$0.5$                        & --                                      \\
    Lagrange point 1
      & $L_{1*}$, $L_{1\circ}$
      & $4.6$--$9.5\,R_{\sun}$ & $2.7$--$6.9\,R_\mathrm{J}$           & $R_\mathrm{J} \simeq 10^{-1}\,R_{\sun}$ \\
    \hline
    Stellar UV flux & $F_\mathrm{UV}$
      & $5\times10^2$--$5\times10^5$ & --               & erg\,cm$^{-2}$\,s$^{-1}$ \\
    Orbital radius  & $R_\mathrm{orb}$
      & --                           & $0.023$--$0.047$ & AU                       \\
    Orbital period  & $\mathcal{P}_\mathrm{orb}$
      & --                           & $1.2$--$3.7$     & days                     \\
    Orbital speed   & $v_\mathrm{orb}$
      & --                           & $138$--$195$     & km\,s$^{-1}$             \\
    \hline
  \end{tabular}
\end{table*}

Table~\ref{tab:parameters} lists the basic model parameters that characterize
the host star and the Jupiter-like close-in planet.
The masses ($M_*$, $M_\circ$) enter only in the gravitational field at the
exterior of each body, whereas their radii ($R_*$, $R_\circ$) define the spheres
within which an internal boundary is applied.
The temperatures correspond to the values at the base of the respective outflows
(the stellar corona and the UV absorption layer in the outer planetary
atmosphere).
The value of the density of the stellar outflow, $\rho_*$, is chosen so that the
mass-loss rate be comparable to that of the solar wind (a few times
$10^{-14}\,M_{\sun}\,\mathrm{yr}^{-1}$), an approach adopted from
\citet{MattPudritz08}.
For the planet, the base density value is determined from the requirement that
the simulated wind match high-resolution 1D numerical simulations that include
the relevant heating and cooling processes (see Sect.~\ref{sec:outflows}).
Thus, $\rho_\circ$ should not be regarded as the actual density at optical depth
$\tau \sim 1$ (the physical base of the outflow) but rather as a numerically
motivated value that gives the appropriate wind profile.
The initial magnetic fields of the star and the planet are assumed to be
dipolar, with equatorial surface values equal to $B_*$ and $B_\circ$,
respectively.

The escape speed from the stellar surface is given by
$v_\mathrm{esc\,*} = (2GM_*/R_*)^{1/2}$, whereas the sound speed at the base of
the stellar wind is $c_\mathrm{s\,*} = (\gamma P_*/\rho_*)^{1/2}$.
A useful model parameter for determining the properties of the wind is
$\lambda_*\equiv (1/2)(v_\mathrm{esc\,*}/c_\mathrm{s\,*})^2$.
Another important nondimensional variable is the ratio of the thermal and
magnetic pressures at the stellar surface, $\beta_* \equiv P_*/(B_*/8\pi)$ (the
stellar plasma beta), which characterizes the dynamical significance of the
magnetic field.\footnote{For plasma-$\beta$ values less than 1, the magnetic
field dominates the dynamics and constrains the plasma to move along flux tubes.
Conversely, when the value of this parameter exceeds 1, the fluid dominates the
dynamics and the magnetic field is dragged by the flow.}
The third relevant parameter for characterizing the outflow is
$\epsilon_* \equiv (\Omega_*R_*/c_{\rm s\,*})^2$, which measures the effect of
surface rotation.
Corresponding expressions for $v_\mathrm{esc\,\circ}$, $c_\mathrm{s\,\circ}$,
$\lambda_\circ$, $\beta_\circ$, and $\epsilon_\circ$ are used for the planet.

The stellar UV flux, $F_\mathrm{UV}$, is a critical quantity because it
determines the energy input and hence the strength of the planetary wind.
Finally, Table~\ref{tab:parameters} gives the rotational periods of the star
($\mathcal{P}_* = 2\pi/\Omega_*$) and the planet
($\mathcal{P}_\circ = 2\pi/\Omega_\circ$) as well as the orbital
characteristics, $R_\mathrm{orb} = x_\circ$,
$\mathcal{P}_\mathrm{orb} = 2\pi/\Omega_\mathrm{orb}$, and
$v_\mathrm{orb} = (GM_*/R_\mathrm{orb})^{1/2}$.
The host star is taken to rotate at $1\%$
of the breakup angular speed (as compared to $0.4\%$ of $(GM_*/R_*^3)^{1/2}$ for
the present sun), whereas the planet is assumed to be tidally locked (i.e.,
$\vec \Omega_\circ = \vec \Omega_\mathrm{orb}$), which is a reasonable
approximation for a close-in planet that interacts tidally with its host star
\citep[e.g.,][]{Matsumura+10}.
Our model planets thus rotate at $\sim$$3$--$25\%$ of breakup.

\subsection{Stellar and planetary winds}

We base the numerical treatment of the two outflows on the simplified approach
of \citet{MattBalick04}.
In brief, we initialize dipolar magnetospheres (force-free by definition) and
keep fixed at each surface the physical quantities that drive the flow.
The temporal evolution of such configurations opens up the magnetospheres and
leads to steady-state winds \citep[see][]{MattPudritz08}.
Although this method does not require the winds to be specified explicitly, we
have opted to initialize a simple form of outflow to ensure that the starting
wind profiles are in agreement with detailed, self-consistent models in the
literature.
We emphasize that, because of the transonic nature of these flows, the final
steady state does not depend on the initial configurations, only on the boundary
conditions that are imposed at the surfaces of the two objects.
However, in low-resolution simulations that do not include all the relevant
physics, simple assumptions about the density and temperature at the boundary
might not give correct results.
Therefore, we use the profiles of detailed outflow models as a guide in order to
set up the appropriate boundary conditions.
This procedure is particularly useful for capturing the correct mass loss rate
of the planet, as described in more detail in Sect.~\ref{sec:outflows}.

We initialize the velocity field of each outflow at time $t = 0$ as an
isotropic, isothermal Parker wind \citep{Parker58}.
In particular, the velocity profile for the stellar outflow,
$v_{\mathrm{w}\,*}^\mathrm{init}(R)$, is obtained by solving numerically the
equation:
\begin{equation}
  \psi_* - \ln\psi_* = - 3 - 4\ln\left(\frac{\lambda_*}{2}\right) + \ln\xi_*
    + 2\frac{\lambda_*}{\xi_*}\,,
  \label{eq:parker}
\end{equation}
where $\psi_*(R) \equiv (v_{\mathrm{w}\,*}^\mathrm{init}/c_\mathrm{s\,*})^2$,
$\xi_*(R) \equiv R/R_*$, and $\lambda_*$ is defined in
Sect.~\ref{sec:parameters}.
In this equation we take the speed of sound to be strictly isothermal, i.e.,
$c_\mathrm{s} = (2k_\mathrm{B}T/m_\mathrm{p})^{1/2}$ (corresponding to a
polytropic index that is exactly 1).
Since the star is assumed to be rotating, we make the approximation that the
initial wind is rotating rigidly with it.
This is a minor contribution to the outflow velocity because $\Omega_*$ is an
order of magnitude smaller than $\Omega_\mathrm{fr}$ (and, in any case, the flow
self-consistently assumes the correct values of $v_\phi$ as the simulation
evolves).
The Cartesian components of the initial stellar outflow velocity are then given
by the sum of the wind, rotation, and frame speeds:
\begin{equation}
  v_{x\,*}^\mathrm{init}(x,y,z) =
    \sin\theta\left[\cos\phi\,v_{\mathrm{w}\,*}^\mathrm{init}(R)
    + \sin\phi\,R(\Omega_\mathrm{fr} + \Omega_*)\right]\hat x\,,
\end{equation}
\begin{equation}
  v_{y\,*}^\mathrm{init}(x,y,z) =
    \sin\theta\left[\sin\phi\,v_{\mathrm{w}\,*}^\mathrm{init}(R)
    - \cos\phi\,R(\Omega_\mathrm{fr} + \Omega_*)\right]\hat y\,,
\end{equation}
\begin{equation}
  v_{z\,*}^\mathrm{init}(x,y,z) =
    \cos\theta\,v_{\mathrm{w}\,*}^\mathrm{init}(R)\,\hat z\,.
\end{equation}
To derive the corresponding expressions for the planetary outflow, we first
calculate $v_{\mathrm{w}\,\circ}^\mathrm{init}(R')$ from Eq.~(\ref{eq:parker}),
using $\psi_\circ(R')$, $\lambda_\circ$, and $\xi_\circ(R') \equiv R'/R_\circ$.
We then add the planet's orbital velocity,
$v_\mathrm{orb} = R_\mathrm{orb}\Omega_\mathrm{orb}\hat y$, as well as the frame
and planetary-rotation velocities (using the rigid-body assumption), to obtain:
\[
  v_{x\,\circ}^\mathrm{init}(x,y,z)
    = \sin\theta'\big[\cos\phi'v_{\mathrm{w}\,\circ}^\mathrm{init}(R')
\]
\begin{equation}
  \quad\quad\quad\quad\quad\quad\quad\quad
    + \sin\phi\,R\Omega_\mathrm{fr} - \sin\phi'R'\Omega_\circ\big]\,\hat x\,,
\end{equation}
\[
  v_{y\,\circ}^\mathrm{init}(x,y,z)
    = \sin\theta'\big[\sin\phi'v_{\mathrm{w}\,\circ}^\mathrm{init}(R')
\]
\begin{equation}
  \quad\quad\quad\quad\quad\quad\quad\quad
    - \cos\phi\,R\Omega_\mathrm{fr} + R_\mathrm{orb}\Omega_\mathrm{orb}
    + \cos\phi'R'\Omega_\circ\big]\,\hat y\,,
\end{equation}
\begin{equation}
  v_{z\,\circ}^\mathrm{init}(x,y,z)
    = \cos\theta'v_{\mathrm{w}\,\circ}^\mathrm{init}(R')\,\hat z\,,
\end{equation}
where $R'$, $\theta'$, and $\phi'$ are functions of $(x,\,y,\,z)$, as explained
at the beginning of this section.
For simplicity, we have assumed that the initial planetary wind orbits and spins
rigidly with the planet, an approximation that is appropriate only close to the
surface.
Note that, for a tidally locked planet, the orbital velocity cancels out with
the frame and planetary rotation velocities, and the planet stays fixed at the
location $(x_\circ,\,y_\circ,\,z_\circ)$.
For example, the $y$ components of these three terms at a distance $x$ along the
line that connects the centers of the star and the planet are
$-x\Omega_\mathrm{fr} + x_\circ\Omega_\mathrm{orb} + x'\Omega_\circ = 0$.
Under this assumption, given that we are working in the corotating frame, we
could have just as well initialized the velocity by using only the thermal
(Parker) wind, with no extra terms.

To obtain the pressure profile of the stellar wind, we solve analytically the
hydrodynamic radial momentum equation,
\begin{equation}
  \rho v_R\frac{dv_R}{dR} + \frac{dP}{dR} = \rho\frac{GM_*}{R^2}\,,
\end{equation}
which gives:
\begin{equation}
  P_*^\mathrm{init}(x,y,z)
    = P_*\exp\left[\lambda_*\left(\frac{R_*}{R} - 1\right) - \frac{1}{2}\left(
    \frac{v_{\mathrm{w}\,*}^\mathrm{init}}{c_\mathrm{s\,*}}\right)^2\right]\,.
\end{equation}
The density is then found using the isothermal-plasma assumption:
\begin{equation}
  \rho_*^\mathrm{init}(x,y,z) = \frac{\rho_*}{P_*}P_*^\mathrm{init}\,.
\end{equation}
In a similar vein, we get for the planet 
\begin{equation}
\label{P_p}
  P_{\circ}^\mathrm{init}(x,y,z) = P_\circ\exp\left[\lambda_\circ\left(
    \frac{R_\circ}{R'} - 1\right) - \frac{1}{2}\left(
    \frac{v_{\mathrm{w}\,\circ}^\mathrm{init}(R')}
    {c_{\mathrm{s}\,\circ}}\right)^2\right]
\end{equation}
and
\begin{equation}
  \rho_{\circ}^\mathrm{init}(x,y,z)
    = \frac{\rho_\circ}{P_\circ}P_\circ^\mathrm{init}\,.
\end{equation}

We set up the stellar wind everywhere in the computational box, apart from a
sphere of radius $10R_\circ$, centered on the planet.
Within this volume we initialize the planetary wind.
Both outflows are initially supersonic and super-Alfv\'enic (with the Alfv\'en
speed given by $v_\mathrm{A} = [B^2/4\pi\rho]^{1/2}$) at the interface that
separates them.\footnote{Recall that in MHD there are three types of waves
(listed in order of increasing propagation speed): slow-magnetosonic (SMS),
Alfv\'en, and fast-magnetosonic (FMS).
Along field lines, an FMS wave propagates at
$\max(v_\mathrm{A},\,c_\mathrm{s})$ and an SMS one at
$\min(v_\mathrm{A},\,c_\mathrm{s})$.
Across field lines, the SMS and Alfv\'en wave propagation speeds are zero, and
an FMS wave propagates at
$(v_\mathrm{A}^2 + c_\mathrm{s}^2)^{1/2}$.
The FMS critical surface is the separatrix beyond which no perturbation can
propagate oppositely to the flow (analogous to the sonic critical surface in
hydrodynamics); we henceforth refer to this surface as the ``critical surface''.
We also follow a common practice in the literature and plot the Alfv\'en and
sonic separatrices rather than the SMS, Alfv\'en, and FMS ones.}

It is worth noting that the Parker wind (Eq.~\ref{eq:parker}) is nondimensional
(with the radial distance expressed in units of the radius of the object and the
velocity in units of the sound speed at the base) and that its solution depends
only on the parameter $\lambda$, which can be written in the form
\begin{equation}
  \lambda = \frac{1}{2}\left(\frac{2GM/R}{2k_\mathrm{B}T/m_\mathrm{p}}\right)
    = \lambda_{\sun}\left(\frac{M}{M_{\sun}}\right)
    \left(\frac{R_{\sun}}{R}\right)\left(\frac{T_{\sun}}{T}\right)\,.
\end{equation}
Interestingly, for a solar analog ($R_* = 1\,R_{\sun}$, $M_* = 1\,M_{\sun}$,
$T_* = 10^6\,\mathrm{K}$) and a Hot Jupiter
($R_\circ = 1\,R_\mathrm{J} \simeq 10^{-1}\,R_{\sun}$,
$M_\circ = 1\,M_\mathrm{J}\simeq 10^{-3}\,M_{\sun}$,
$T_\circ = 10^4\,\mathrm{K}$) we obtain $\lambda_* \approx \lambda_\circ$.
As a result, the two outflows reach their corresponding sonic speeds at the same
scaled distance.
For the adopted fiducial parameters, $\lambda = 11.5$, which implies that the
stellar and planetary outflows become supersonic at a radius that is slightly
smaller than $6R_*$ and $6R_\circ$, respectively.

In a real system, the stellar flux that heats the planetary surface varies with
longitude (in particular, there is no direct irradiation of the planet's night
side) as well as latitude (in particular, the planetary poles also receive no
direct heating).
This implies that even an unmagnetized outflow would be anisotropic, although
the quantitative effect of the uneven irradiation on the evaporative mass
outflow rate need not be large \citep[e.g.,][]{MurrayClay+09}.
In this work we simplify the treatment by assuming that the base temperature and
density are uniform over the entire planetary surface.

The initial magnetic field configuration in the computational domain is taken to
be a sum of two magnetic dipoles,
$\vec B_*^\mathrm{init} + \vec B_\circ^\mathrm{init}$, which are given by
\begin{equation}
 \vec B_*^\mathrm{init}(x,y,z) = \frac{B_*R_*^3}{R^5}
   \left[3xz\,\hat x + 3yz\,\hat y + \left(3z^2-R^2\right)\,\hat z\right]
\end{equation}
and
\begin{equation}
 \vec B_\circ^\mathrm{init}(x,y,z) = \frac{B_\circ R_\circ^3}{R'^5}
   \left[3x'z'\,\hat x + 3y'z'\,\hat y
   + \left(3z'^2-R'^2\right)\,\hat z\right]
\end{equation}
for the star and planet, respectively.
Both dipoles are oriented along the $z$ axis and are aligned with each other.
We also write explicitly the components of the total gravitational field
$\vec g$ outside the two bodies:
\begin{equation}
  \vec g_*(x,y,z) = \frac{GM_*}{R^3}\vec R
    = \frac{GM_*}{R^3}\left(x\,\hat x + y\,\hat y + z\,\hat z\right)\, ,
\end{equation}
\begin{equation}
  \vec g_\circ(x,y,z)
    = \frac{GM_\circ}{R'^3}\left(x'\,\hat x + y'\,\hat y + z'\,\hat z\right)\,.
\end{equation}

\subsection{Wind boundary conditions}

In order to continuously accelerate the outflows and guarantee a constant supply
of mass, we keep fixed the density, pressure, and velocity at their bases.
Specifically, during the temporal evolution, we impose the initial values of
these quantities at the regions defined by $R_* < R < 1.5\,R_*$ and
$R_\circ < R' < 1.5\,R_\circ$ for the stellar and planetary winds, respectively.
However, the magnetic field is free to evolve within these shells, readjusting
self-consistently from its initial dipolar topology.
This is a simplified treatment as compared with the formulation of
\cite{MattPudritz08}, who modeled 2.5D axisymmetric stellar winds using a
thinner, four-layer shell above the surface of the star.\footnote{A 2.5D MHD
simulation refers to the evolution of all three components of the velocity and
magnetic field in a 2D computational domain.}
Within that region, they progressively relaxed the time-dependency constraint on
the physical quantities.
Here we employ a less detailed setup on account of the lower resolution imposed
by our 3D modeling.
Nevertheless, the steady-state winds that we obtain compare well with the more
refined 2.5D simulations (see Sect.~\ref{sec:outflows}).

The above implementation of stellar winds differs from the approach adopted by
\cite{Cohen+11a, Cohen+11b, Cohen+14} and \cite{Vidotto+14} (and references
therein), which includes nonaxisymmetric and/or temporal variabilities.
We neglect such effects in this work in order to investigate the star--planet
interaction in a general manner that does not depend on temporary local
fluctuations.

Our wind boundary conditions do not incorporate the effect of tidal forces on
the planetary outflow.
As discussed by \citet{Trammell+11}, these forces could suppress the wind from
the polar regions of a tidally locked planet with a dipolar field geometry.
In particular, these authors showed that the wind would not undergo a sonic
transition in this case if $\epsilon_\circ > 4/\lambda_\circ^2$ (where the
parameters $\epsilon$ and $\lambda$ are defined in Sect.~\ref{sec:parameters}).
Our neglect of this suppression in the adopted boundary conditions is not a
serious omission in view of the fact that our simulations incorporate the effect
of these forces for $R'\ge1.5\,R_\circ$ and that the sonic surface typically
lies at a distance of a few $R_\circ$ from the planet's center.\footnote{ We
note in this connection that the only two models in Table~\ref{tab:models} that
develop supercritical outflows, \texttt{FvRb} and \texttt{FvRB}, do \emph{not}
satisfy the inequality $\epsilon_\circ > 4/\lambda_\circ^2$.
This is consistent with the fact that the sonic surface plotted in
Fig.~\ref{fig:FvRb} is represented by a \emph{closed} curve.}
Furthermore, for the star--planet configurations considered in this paper, which
are characterized by parallel orbital and spin axes, the bulk of the interaction
between the stellar and the planetary plasmas occurs in the equatorial plane,
with the details of the planetary outflow from the polar regions having little
effect on the results.

\subsection{Stellar and planetary interiors}

The interiors of both the star and the planet are treated as an internal
boundary.
In particular, the values of all physical quantities are kept fixed by having
them overwritten at each time step.
To avoid spurious effects at the surfaces of the objects, or extreme dynamics
that could affect the global time step of the simulation, we prescribe an
approximate equilibrium within their volumes.

Since the physical conditions in the interior of the star or the planet are not
important for the simulation, the simplest approach is to consider them as
uniform-density spheres.
The interior gravitational accelerations are then given by
\begin{equation}
  \vec g_*(x,y,z) = \frac{4}{3}\pi G \rho_*\vec R
    = \frac{4}{3}\pi G \rho_* (x\,\hat x + y\,\hat y + z\,\hat z)
\end{equation}
and
\begin{equation}
  \vec g_\circ(x,y,z)
    = \frac{4}{3}\pi G \rho_\circ (x'\hat x + y'\hat y + z'\hat z)\,,
\end{equation}
from which the pressures can be inferred using the hydrostatic equilibrium
condition:
\begin{equation}
  P_*^\mathrm{init}(x,y,z)
    = P_* + \frac{2}{3}\pi G\rho_*^2\left(R_*^2 - R^2\right)\,,
\end{equation}
\begin{equation}
  P_\circ^\mathrm{init}(x,y,z)
    = P_\circ + \frac{2}{3}\pi G\rho_\circ^2\left(R_\circ^2 - R'^2\right)\,.
\end{equation}

We keep the dipolar expression for the magnetic field, but we smooth it out at
the center where it becomes singular.
Specifically, we assume that the central regions of the star and the planet,
$R < R_*/2$ and $R' < R_\circ/2$, are uniformly magnetized spheres, with
magnetic fields $\vec B_*^\mathrm{init} = 16B_*\,\hat z$ and
$\vec B_\circ^\mathrm{init} = 16B_\circ\,\hat z$ that connect smoothly with the
dipolar configurations at larger radii.
Finally, we approximate the star and the planet as rotating rigid bodies and
specify their angular velocities by $\vec\Omega_* = \Omega_*\hat z$ and
$\vec\Omega_\circ = \Omega_\circ\hat z$, respectively.

\subsection{The models}
  \label{subsec:models}

\begin{table*}
  \caption{
    Parameters of the numerical models.
    From left to right: orbital radius ($R_\mathrm{orb}$), planetary radius
    ($R_\circ$) and mass ($M_\circ$), surface escape speed
    ($v_{\rm esc\, \circ}$), distance to the $L_1$ point (in units of
    $R_\circ$), density and ($\rho_\circ$) temperature ($T_\circ$) at the base
    of the outflow, equatorial surface magnetic field ($B_\circ$), and
    irradiating flux (high or low; see text for the numerical values).
    The labels of the models are divided into four parts: the first denotes the
    stellar UV flux (high [\texttt{F}] or low [\texttt{f}]), the next describes
    the magnitude of the escape speed (large [\texttt{V}] or small
    [\texttt{v}]), then its distance from the host star (far [\texttt{R}] or
    near [\texttt{r}]), and finally the strength of its magnetic field (strong
    [\texttt{B}] or weak [\texttt{b}]). 
  \label{tab:models}}
  \centering 
  \begin{tabular}{lcccccccccc}
    \hline
    \hline
    Model & $R_\mathrm{orb}\ (\mathrm{AU})$ & $R_\circ\ (R_\mathrm{J})$
      & $M_\circ\ (M_\mathrm{J})$
      & $v_{\mathrm{esc}\,\circ}\ (\mathrm{km\,s}^{-1})$
      & $L_{1\circ}\ (R_\circ)$ & $\rho_\circ\ (\mathrm{g\,cm}^{-3})$
      & $T_\circ\ (\mathrm{K})$ & $B_\circ\ (G)$ & Irradiation \\
    \hline
    \texttt{FvRB} & $0.047$ & $1.5$ & $0.5$ & $3$5 & $3.7$ & $7\times10^{-16}$ & $10^4$        & $1$   & High UV flux \\
    \texttt{FvRb} & $0.047$ & $1.5$ & $0.5$ & $35$ & $3.7$ & $7\times10^{-16}$ & $10^4$        & $0.1$ & High UV flux \\
    \texttt{FvrB} & $0.023$ & $1.5$ & $0.5$ & $35$ & $1.8$ & $7\times10^{-16}$ & $10^4$        & $1$   & High UV flux \\
    \hline
    \texttt{fvRB} & $0.047$ & $1.5$ & $0.5$ & $35$ & $3.7$ & $7\times10^{-17}$ & $6\times10^3$ & $1$   & Low UV flux  \\
    \texttt{fvRb} & $0.047$ & $1.5$ & $0.5$ & $35$ & $3.7$ & $7\times10^{-17}$ & $6\times10^3$ & $0.1$ & Low UV flux  \\
    \texttt{fvrB} & $0.023$ & $1.5$ & $0.5$ & $35$ & $1.8$ & $7\times10^{-17}$ & $6\times10^3$ & $1$   & Low UV flux  \\
    \hline
    \texttt{FVRB} & $0.047$ & $1$   & $1$   & $60$ & $6.9$ & $10^{-13}$        & $10^4$        & $1$   & High UV flux \\
    \texttt{FVRb} & $0.047$ & $1$   & $1$   & $60$ & $6.9$ & $10^{-13}$        & $10^4$        & $0.1$ & High UV flux \\
    \texttt{FVrB} & $0.023$ & $1$   & $1$   & $60$ & $3.4$ & $10^{-13}$        & $10^4$        & $1$   & High UV flux \\
    \hline
    \texttt{fVRB} & $0.047$ & $1$   & $1$   & $60$ & $6.9$ & $3\times10^{-13}$ & $6\times10^3$ & $1$   & Low UV flux  \\
    \texttt{fVRb} & $0.047$ & $1$   & $1$   & $60$ & $6.9$ & $3\times10^{-13}$ & $6\times10^3$ & $0.1$ & Low UV flux  \\
    \texttt{fVrB} & $0.023$ & $1$   & $1$   & $60$ & $3.4$ & $3\times10^{-13}$ & $6\times10^3$ & $1$   & Low UV flux  \\
    \hline
  \end{tabular}
\end{table*}

Table~\ref{tab:models} lists the models that we study numerically.
We aim to explore the observationally relevant regions of parameter space and
attempt to deduce, based on the behaviors exhibited by the simulated systems, a
general --- and largely model independent --- classification scheme for
magnetized star--planet interactions.
We usually adopt two representative values for each of the physical variables
that need to be specified in the model.
Thus, we consider two combinations of planetary parameters, the first having
Jupiter's mass and radius and the other corresponding to a less massive
($0.5\,M_\mathrm{J}$) and larger-radius ($1.5\,R_\mathrm{J}$) planet.
These choices are intended to sample the typical range of values for the escape
speed $v_\mathrm{esc\,\circ} \propto (M_\circ/R_\circ)^{1/2}$ from a hot
Jupiter, which is relevant to the mass evaporation rate $\dot M_\mathrm{ev}$ in
the low-$F_\mathrm{UV}$ limit (where
$\dot M_\mathrm{ev} \propto 1/v_\mathrm{esc\,\circ}^2$; e.g.,
\citealt{MurrayClay+09}).

We similarly consider two magnitudes for the UV flux, one low
($F_\mathrm{UV} = 5\times 10^2\,\mathrm{erg\,cm}^{-2}\,\mathrm{s}^{-1}$) and one
high ($F_\mathrm{UV}=5\times 10^5\,\mathrm{erg\,cm}^{-2}\,\mathrm{s}^{-1}$),
adopting numerical values similar to those given for these two limits in
\citet{MurrayClay+09}.
These two values correspond to different evaporation regimes.
In the high-flux case, the excess photoionization energy is largely lost by
radiative cooling (radiation/recombination-limited evaporation), whereas in the
low-flux case it is mostly absorbed and drives the outflow (energy-limited
evaporation).
The outflow from a strongly irradiated planet is denser, hotter, and faster than
that from a weakly irradiated one \citep[e.g.,][]{MurrayClay+09}. 
Since in this paper we do not explicitly model the radiative processes in the
planetary atmosphere, we specify the wind properties through the values of the
base temperature and density, which are adjusted to match the results of more
detailed calculations (see Sect.~\ref{sec:outflows}).

We also vary the distance of the Hot Jupiter from the host star.
Smaller orbital radii imply a stronger tidal force as well as a locally weaker,
and possibly not fully developed, stellar wind.\footnote{We do not take into
account changes to the incident UV flux that result from variations in the
orbital radius since their magnitudes remain small in comparison with the
difference between our chosen representative values.}
The value of the orbital radius also determines whether the planet lies within
the critical surface of the stellar wind or whether it is impacted by a
super-critical flow.

The two values that we adopt for the magnetic field amplitude at the planet's
surface differ by an order of magnitude and are intended to represent a
``strong'' and a ``weak'' field.
Stronger fields correspond to higher magnetic pressure and tension and exhibit a
greater ``rigidity.''
In this case the field resists being opened up by the stellar outflow over a
larger region near the equator, and as a result the planetary magnetosphere
presents a larger obstacle in its interaction with the stellar wind.

Although we vary the stellar UV flux, our simulations employ only a single set
of stellar wind parameters.
Our coverage of the relevant parameter phase space is, however, sufficiently
broad to yield general results, which in principle apply also to systems with
either a weaker or a stronger stellar wind.

\subsection{Numerical setup}

The simulations are performed with PLUTO (version 4.0.1), a code for
computational astrophysics \citep{Mignone+07, Mignone+12}.\footnote{PLUTO is
freely available at \texttt{http://plutocode.ph.unito.it}}
The computational domain consists of a cube, with $x,y,z\in[-15,\,15]\,R_*$,
which is resolved by a static, multiply refined grid of $224\times192\times192$.
We have also carried out one simulation (model \texttt{FvRb}) in higher
resolution, $424\times320\times320$, in order to validate our results.
The region around the star, $x,y,z\in[-5,\,5]\,R_*$, is resolved with
$\Delta x,\, \Delta y,\, \Delta z \simeq 0.16\,R_*$ ($64^3$ cells) in most cases,
and with $\Delta x,\, \Delta y,\, \Delta z \simeq 0.08\,R_*$ ($128^3$ cells) in
the high-resolution simulation.
However, for the Hot Jupiter and its surroundings,
$x',y',z'\in[-5,\,5]\,R_\circ$, we use a resolution that is higher by one order
of magnitude, i.e., $\Delta x,\, \Delta y,\, \Delta z \simeq 0.016\,R_*$ ($64^3$
cells) and $\Delta x,\, \Delta y,\, \Delta z \simeq 0.008\,R_*$ ($128^3$ cells)
for the standard and highly resolved cases, respectively.
In the region between the star and the planet we apply the resolution
$\Delta x, \Delta y, \Delta z \simeq 0.16\,R_*$.

The total time of the simulation is typically $\sim$$4$ days, only a fraction
of which is required for the system to attain a quasi-steady state.
For comparison, a stellar wind propagating at $300\,\mathrm{km\,s}^{-1}$ travels
$\sim$$10$ times the distance from the star to the outer boundary over this time
interval.
We adopt highly accurate numerical schemes to compensate for the resolution
limits imposed by the three-dimensional character of the simulations.
In particular, we use the HLLD Riemmann solver and apply third-order accuracy in
space (piecewise parabolic interpolation) and time ($3^\mathrm{rd}$ order
Runge-Kutta).
The condition $\nabla\cdot\vec B = 0$ is implemented using the
divergence-cleaning method, an approach based on the generalized Lagrange
multiplier (GLM) formulation (see \citealt{Mignone+12} and references therein).

\subsection{Verification and calibration of 3D wind models}
  \label{sec:outflows}

The acceleration and final steady-state properties of a simulated wind depend on
the included physics as well as on the resolution of the grid.
3D simulations are computationally expensive and therefore cannot incorporate
the same level of detail that is possible in 1D and 2D models.
In this work we have implemented a simplified procedure for treating the stellar
and planetary outflows, and we now check on the validity of the adopted
approximations.
We first compare our 3D stellar-wind model with a more detailed (and
higher-resolution) 2.5D model and verify its consistency.
We then demonstrate our method of ensuring the consistency of 3D planetary wind
models with 1D hydrodynamic calculations that incorporate the relevant physics
of evaporative outflows.

\begin{figure}
  \resizebox{\hsize}{!}{\includegraphics{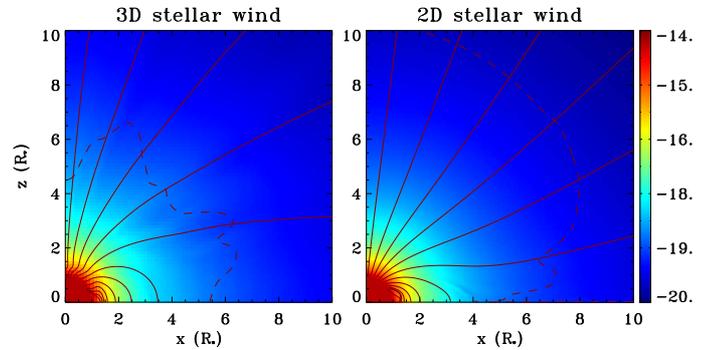}}
  \caption{
    Logarithmic density contours, field lines (solid lines), and poloidal
    Alfv\'en surface (dashed line) in the final steady state of the 3D stellar
    wind model employed in this paper (left panel) and of a 2.5D high-resolution
    simulation (right panel).
  \label{fig:stellarwind}}
\end{figure}
Figure~\ref{fig:stellarwind} compares the 3D stellar wind model adopted in this
work (left panel), with a 2.5D simulation with the same parameters (right
panel), but with a $10$ times higher resolution along each direction
\citep[modeled as in][]{MattPudritz08}.
The density, velocity, and magnetic field have very similar profiles, despite
the fact that every cell on the left would be resolved by a thousand cells on
the right if it were a 3D setup.
Any discrepancies can be neglected since the outflow serves its purpose of
providing a generic stellar wind.

\begin{figure}
  \resizebox{\hsize}{!}{\includegraphics{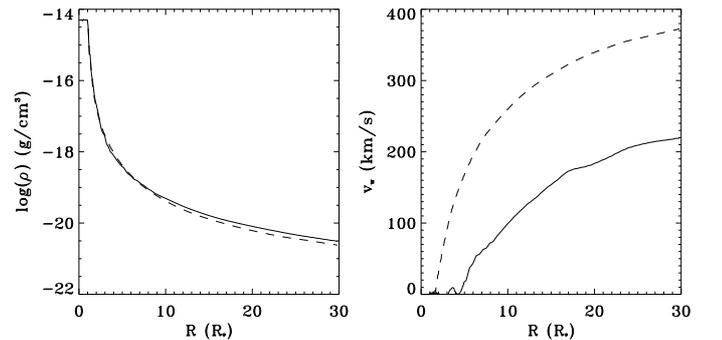}}
  \caption{
    Logarithmic density (left) and radial velocity (right) in the final steady
    state of the 3D stellar wind model as a function of the distance $R$ from
    the center of the star along the equatorial ($x$) and axial ($z$) directions
    (solid and dashed lines, respectively).
  \label{fig:stellarwindprofile}}
\end{figure}
Figure~\ref{fig:stellarwindprofile} shows the profiles of the density (left
panel) and radial velocity (right panel) for the steady state reached by our
stellar wind model.
The radial velocity is zero along the equator up to $\sim$$4R_*$, corresponding
to the extent of the stellar dead zone (see Fig.~\ref{fig:stellarwind}).
More generally, the finding that the wind speed remains lower along $x$ than
along $z$ for any given value of $R$ can be attributed to the fact that a
smaller fraction of the field lines open up near the equator than near the
poles.
This behavior is also exhibited by more detailed models of winds from solar-type
stars \citep[e.g.,][]{CohenDrake14}.
In the case of  faster rotators, in which magnetocentrifugal effects play a more
prominent role in the acceleration of the wind, the outflow speed along the
equator can become higher than along the spin axis \cite[e.g.,][]
{MattPudritz08}.
This regime is pertinent to our models of tidally locked planets.

\begin{figure}
  \resizebox{\hsize}{!}{\includegraphics{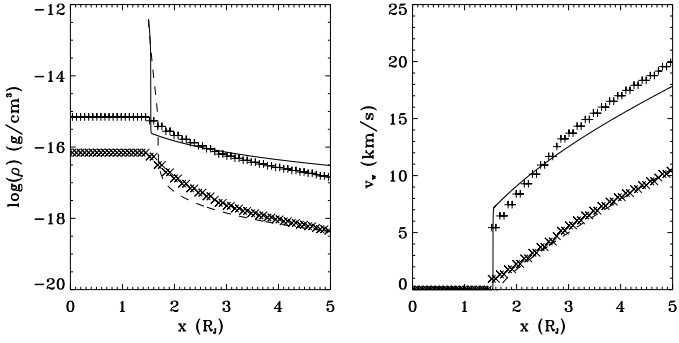}}
  \caption{
    Logarithmic density (left) and radial velocity (right) as a function of
    distance from the planet's center for the isotropic outflow model used to
    initialize the 3D wind simulation for a ``light'' Hot Jupiter
    ($M_\circ = 0.5\,M_\mathrm{J}$, $R_\circ = 1.5\,R_\mathrm{J}$).
    The symbols ``$+$'' and ``$\times$'' denote, respectively, the high- and
    low-flux cases (modeled with $T_\circ = 10^4\,\mathrm{K}$ and
    $T_\circ = 6\times10^3\,\mathrm{K}$, respectively).
    The solid and dashed lines are from the corresponding 1D simulations that
    take into account the relevant heating and cooling processes (see text for
    details).
  \label{fig:planet1wind}}
\end{figure}
\begin{figure}
  \resizebox{\hsize}{!}{\includegraphics{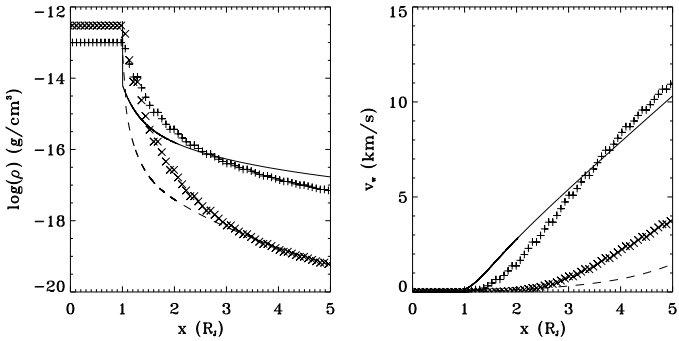}}
  \caption{
    Same as Fig.~\ref{fig:planet1wind}, but for a Jupiter-like planet
    ($M_\circ = M_\mathrm{J}$, $R_\circ = R_\mathrm{J}$).
  \label{fig:planet2wind}}
\end{figure}
To ensure that the planetary winds are properly initialized, we specify the
boundary conditions so that the resulting profiles match the results of a more
elaborate (and highly resolved) 1D model that incorporates relevant additional
physics.
Specifically, we simulate the outflow along the line joining the centers of the
planet and the star, incorporating explicit energy and ionization balance
equations following the formulation of \citet{MurrayClay+09}.
The energy equation includes advective and $PdV$-work terms as well as source
terms in the form of photoionization heating (assuming a single ionizing photon
energy of $20\,$eV and perfect efficiency) and Ly$\alpha$ cooling (under the
assumption of collisional excitation by free electrons).
This equation is integrated using a modified version of the Simplified
Non-Equilibrium Cooling (SNEq) module of PLUTO \citep{Tesileanu+08}.
The hydrogen ionization balance equation accounts for photoionization, Case B
radiative recombination, and ion advection.
Using the results of these calculations, we fine-tuned the boundary conditions
of the simplified 3D models until the latter were found to effectively capture
the general features of the 1D models} for our representative planets in both the
high- and low-flux limits.
The steady-state planetary wind profiles obtained in this way are shown in
Figs.~\ref{fig:planet1wind} and~\ref{fig:planet2wind} together with the
corresponding 1D results.

As can be seen from Figs.~\ref{fig:planet1wind} and~\ref{fig:planet2wind}, the
1D outflow models predict a very steep density drop near the surface of the
planet.
This implies that, if one were to choose the value of the surface density
$\rho_\circ$ for the 3D simulation simply on the basis of the location of the
nominal base of the flow (where $\tau\sim 1$), this would likely not lead to a
good match with the 1D density profile.
On the other hand, if one were to choose even a slightly different reference
radius for fixing the surface density, this might result in a significant under-
or over-estimate of the mass outflow rate.
These considerations strongly suggest that the correct procedure for modeling
the underlying planetary outflow is to choose the boundary values of density and
temperature on the surface of the planet so that they lead to a good match with
the reference profiles.
The small discrepancy seen in the velocity profile of the low-flux case in
Fig.~\ref{fig:planet2wind} can be ignored because, as we show below, the
interaction with the stellar plasma chokes the outflow from a Jupiter-like
planet for this value of the flux.

\section{Numerical results}
  \label{sec:results}

\subsection{Unmagnetized star--planet interactions}

Before turning to the MHD simulations, we make a few qualitative remarks on the
expected behavior in the absence of magnetic fields.
Assuming that there is no planetary outflow, the stellar wind will impact the
planet directly.
For a supersonic stellar flow, the height above the planet's surface where the
plasma will get shocked is determined by the location where the atmospheric
thermal pressure, $P_{\mathrm{th}\,\circ}$, is equal to the local value of the
wind ram pressure, $\rho_{\mathrm{w}\,*}v_{\mathrm{w}\,*}^2$.
For a  subsonic flow, there will be no shock, but rather a smooth transition of
the physical quantities between the two bodies: the density and pressure will
have high values at the surface of each object and will be lower in between.
The converse will hold for the velocity: zero values at the surfaces, and a
smooth acceleration --- and subsequent deceleration --- on moving from the host
to the Hot Jupiter.

If a planetary outflow is present, its strength will determine whether it will
be suppressed by the stellar wind (weak outflow) or else become supersonic and
then collide with the stellar plasma well away from the planet's surface (strong
outflow).
When both flows are supersonic, they interact at the location where the ram
pressure of the planetary wind, $\rho_{\mathrm{w}\,*}v_{\mathrm{w}\,*}^2$, is
equal to that of the stellar gas (as measured in the corotating frame).

\subsection{General behavior}

We start our presentation of the MHD simulations by describing the general
structure of the interacting outflows; in the ensuing subsections we present the
results for each of the simulated models in greater detail.
\begin{figure}
  \resizebox{\hsize}{!}{\includegraphics{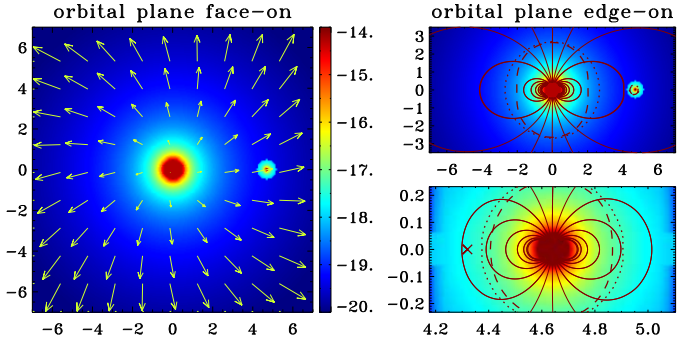}}
  \caption{
    Initial conditions in the lab frame for model \texttt{FVRB}.
    The axes are in units of $0.01\,\mathrm{AU}$.
    Logarithmic density contours (in units of $\mathrm{g\,cm}^{-3}$) are shown
    in the $x$-$y$ plane ($z=0$; left panel) and in the $x$-$z$ plane ($y = 0$;
    right panels).
    The top right panel is focused on the star and the bottom right on the
    planet.
    Solid lines represent the magnetic field, arrows show the velocity field
    (with the longest one corresponding to $\sim$$290\,\mathrm{km\,s}^{-1}$),
    dashed lines the poloidal Alfv\'en surface
    ($v_x^2 + v_z^2 = (B_x^2 + B_z^2)/4\pi\rho$), and dotted lines the poloidal
    sonic surface ($v_x^2 + v_z^2 = \gamma P/\rho$).
    The cross in the bottom right panel denotes the location of the $L_1$ point.
  \label{fig:initial}}
\end{figure}
We use model \texttt{FVRB} for this illustration, and show its initial
configuration in Fig.~\ref{fig:initial}; the initial structure of the other
models is similar except for the planetary wind profiles, which depend on the
choice of surface parameters (see Figs.~\ref{fig:planet1wind} and
\ref{fig:planet2wind}).
The star is located at the center of the left panel and of the top right panel,
which respectively provide face-on and edge-on views of the orbital plane.
The stellar wind is depicted with arrows, and its initially dipolar magnetic
field with solid lines.
Next to the star, at $x \sim 4.5$ (in units of $0.01\,\mathrm{AU}$), is the
spherical volume within which the Hot Jupiter and its outflow are located (left
and top right panels).
A close-up of the region near the planet is shown in the bottom right panel.
Note that both the stellar and the plentary winds are super-Alfv\'enic and
supersonic at the initial interface between them (a sphere of radius
$10R_\circ$, centered on the planet).

At the beginning of the simulation, the plasma escaping from the surface of each
object forces the magnetosphere to open.
In general, depending on the strength of the magnetic field and the physical
conditions at the base of the wind, this might happen over the entire surface
(for a weaker field and/or a stronger outflow) or just at the polar regions (for
a stronger field and/or a weaker outflow).
For the adopted value of the stellar surface magnetic field
($B_* = 2\,\mathrm{G}$), the star forms a dead zone for
$\pi/4 \lesssim \theta \lesssim 3\pi/4$.
In this region, the hot plasma cannot overcome the opposing forces of stellar
gravity and magnetic tension, both of which are almost perpendicular to the
flow.
The behavior of the planetary magnetosphere will be discussed on a case by case
basis: its structure depends both on $B_\circ$ and the planetary wind properties
\emph{and} on its environment.

\subsection{Models exhibiting a planetary tail}
  \label{sec:tail}

\begin{figure}
  \resizebox{\hsize}{!}{\includegraphics{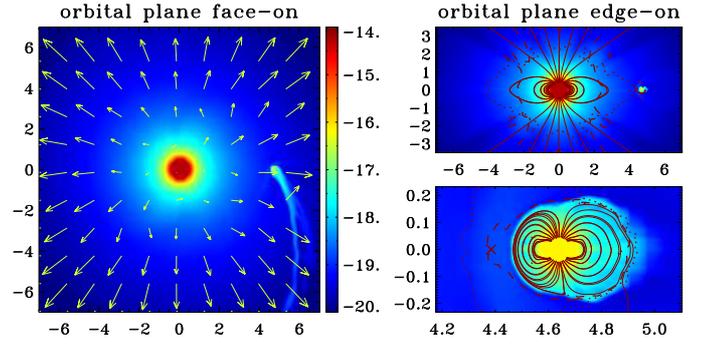}}
  \caption{
    Logarthimic density (color contours in units of g\,cm$^{-3}$), velocity
    (arrows), magnetic field (solid lines), critical surfaces (dotted line:
    sonic, dashed line: Alfv\'en), and the $L_1$ point (cross) for the final
    steady state of model \texttt{fvRb}.
    Distances are measured in units of $0.01\,\mathrm{AU}$.
    The panels are arranged as in Fig.~\ref{fig:initial}, with the left and top
    right panels centered on the star, and the bottom right one on the planet.
  \label{fig:fvRb}}
\end{figure}
Figure~\ref{fig:fvRb} displays the final steady state of model \texttt{fvRb}.
This case corresponds to a low UV flux, and the planetary outflow is weak and
cannot overcome the ram pressure of the stellar wind (projected along the
orbit).
The stellar wind is stopped by the magnetic pressure of the planet's closed
field lines (which exceeds the thermal pressure above the planet's surface),
terminating in a bow shock that stands off a few planetary radii from the
planet's surface.
The magnetosphere is dragged backward by the stellar wind, forming a
cometary-type tail.
This process is akin to the formation of the geomagnetic tail in the interaction
of the solar wind with Earth's magnetosphere \citep[e.g.,][]{Ness65}.
The tail does not follow the trajectory of the planet since it is continuously
pushed away radially by the stellar wind, resulting in a tilted structure.

\begin{figure}
  \resizebox{\hsize}{!}{\includegraphics{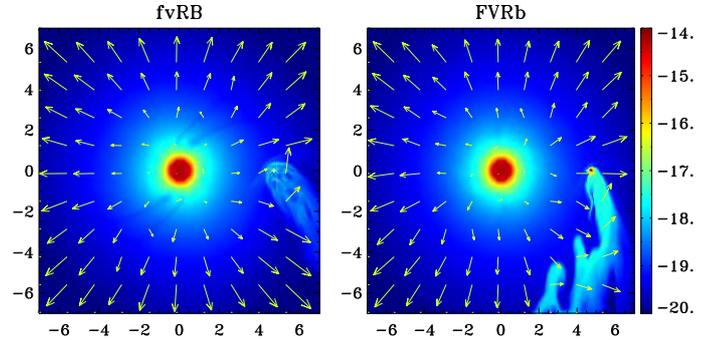}}
  \caption{
    Same as the left panel of Fig.~\ref{fig:fvRb}, but for models \texttt{fvRB}
    (left) and \texttt{FVRb} (right).
  \label{fig:fvRB-FVRb}}
\end{figure}
Figure~\ref{fig:fvRB-FVRb} illustrates how the structure of the tail is
modified when the model parameters are changed.
The left panel shows the results for model \texttt{fvRB}, which has the same
parameters as model \texttt{fvRb} (Fig.~\ref{fig:fvRb}) except that the
planetary magnetic field is a factor of $10$ stronger, corresponding to an
increase by a factor of $100$ in the magnetic pressure at a given distance from
the planet.
As expected, this results in the bow shock being located farther away from the
planetary surface, which, in turn, gives rise to a wider tail.
The right panel presents the final state for model \texttt{FVRb}, which
corresponds to a higher UV flux (which promotes an outflow) but also a larger
escape speed (which hinders a wind).
Overall, its behavior is very similar to that of model \texttt{fvRb}, although
the tail in this case is noticeably denser.
This can be understood from the fact that, in the high-flux limit, the base
density scales roughy as $F_\mathrm{UV}^{1/2}$ \citep{MurrayClay+09}. 

The snapshot shown in the right panel of Fig.~\ref{fig:fvRB-FVRb}
highlights a few interesting dynamical features of cometary-type tails.
The plasma that trails the planet has a velocity comparable to the orbital
speed, whereas the component of the stellar wind along the orbit is negligible
at that location.
This results in the development of strong shear, which may trigger a
Kelvin-Helmholtz instability.
Furthermore, the stellar plasma that pushes the tail outward is of lower density
than the trailing stream, which can induce a Rayleigh-Taylor instability. 
In this particular simulation, their effect on the tail structure remains
comparatively mild, likely on account of the relatively high density of the
trailing stream.
We have, however, found that the induced distortions can be more persistent in
other cases.

\begin{figure}
  \resizebox{\hsize}{!}{\includegraphics{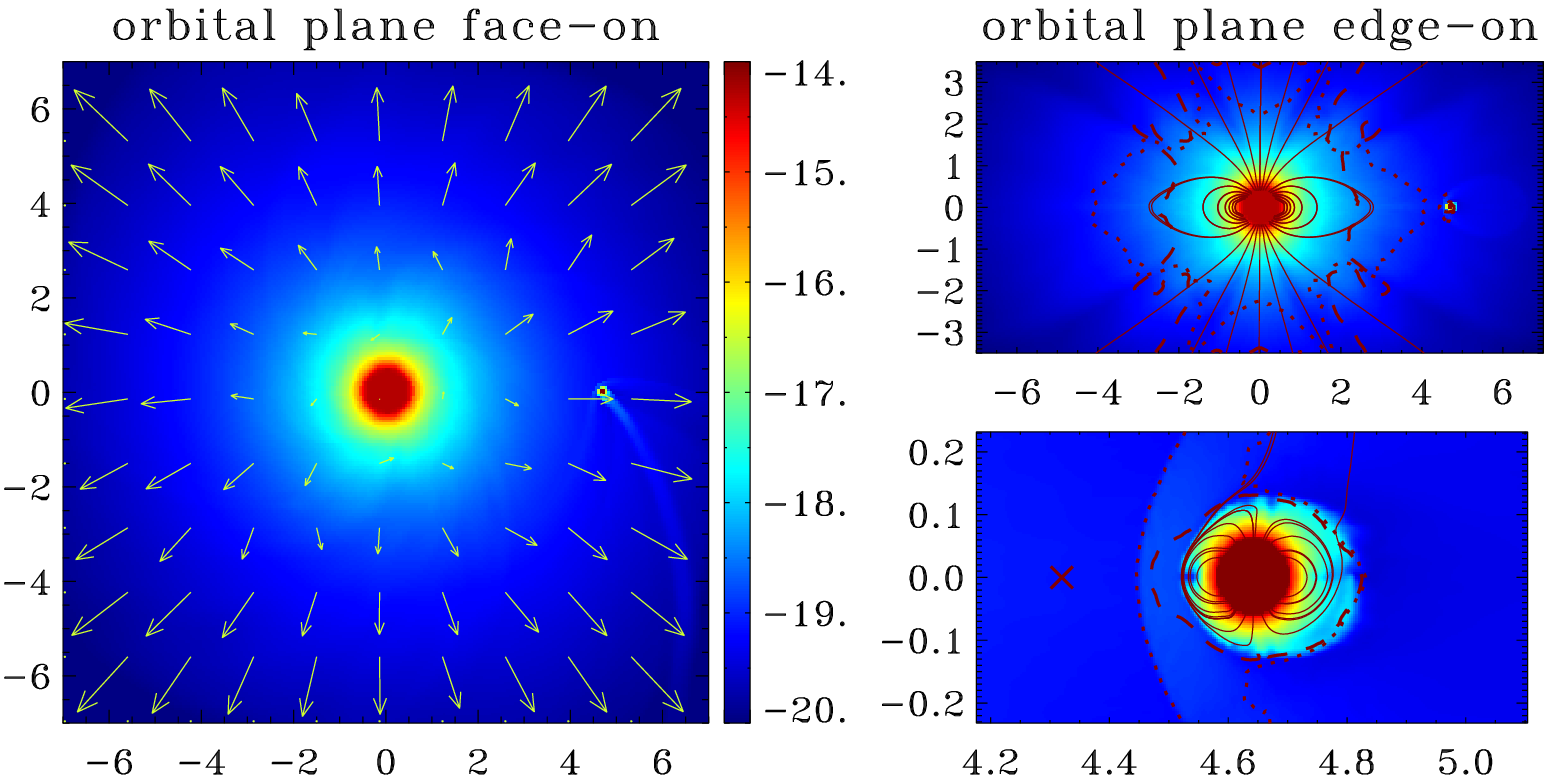}}
  \caption{
    Same as Fig.~\ref{fig:fvRb}, but for model \texttt{fVRb}.
  \label{fig:fVRb}}
\end{figure}
When the UV flux remains low but the escape speed is high, the planetary outflow
weakens and the density in the tail becomes measurably lower than in the models
considered so far.
This result follows directly from the expression for the mass evaporation rate
in the low-flux (energy-limited) limit, which implies that
$\dot M_\mathrm{ev} \propto F_\mathrm{UV}/v_\mathrm{esc}^2$ \citep[see Eq.~9 in]
[]{MurrayClay+09}.
In this case the tail becomes thin and barely noticeable.
This is demonstrated in Fig.~\ref{fig:fVRb} for model \texttt{fVRb} --- the
outflow is even weaker when the magnetic field amplitude is increased
(model \texttt{fVRB}, not plotted).

\subsection{Models exhibiting both a tail and an accretion stream}
\label{sec:collide}

\begin{figure*}
  \resizebox{\hsize}{!}{\includegraphics{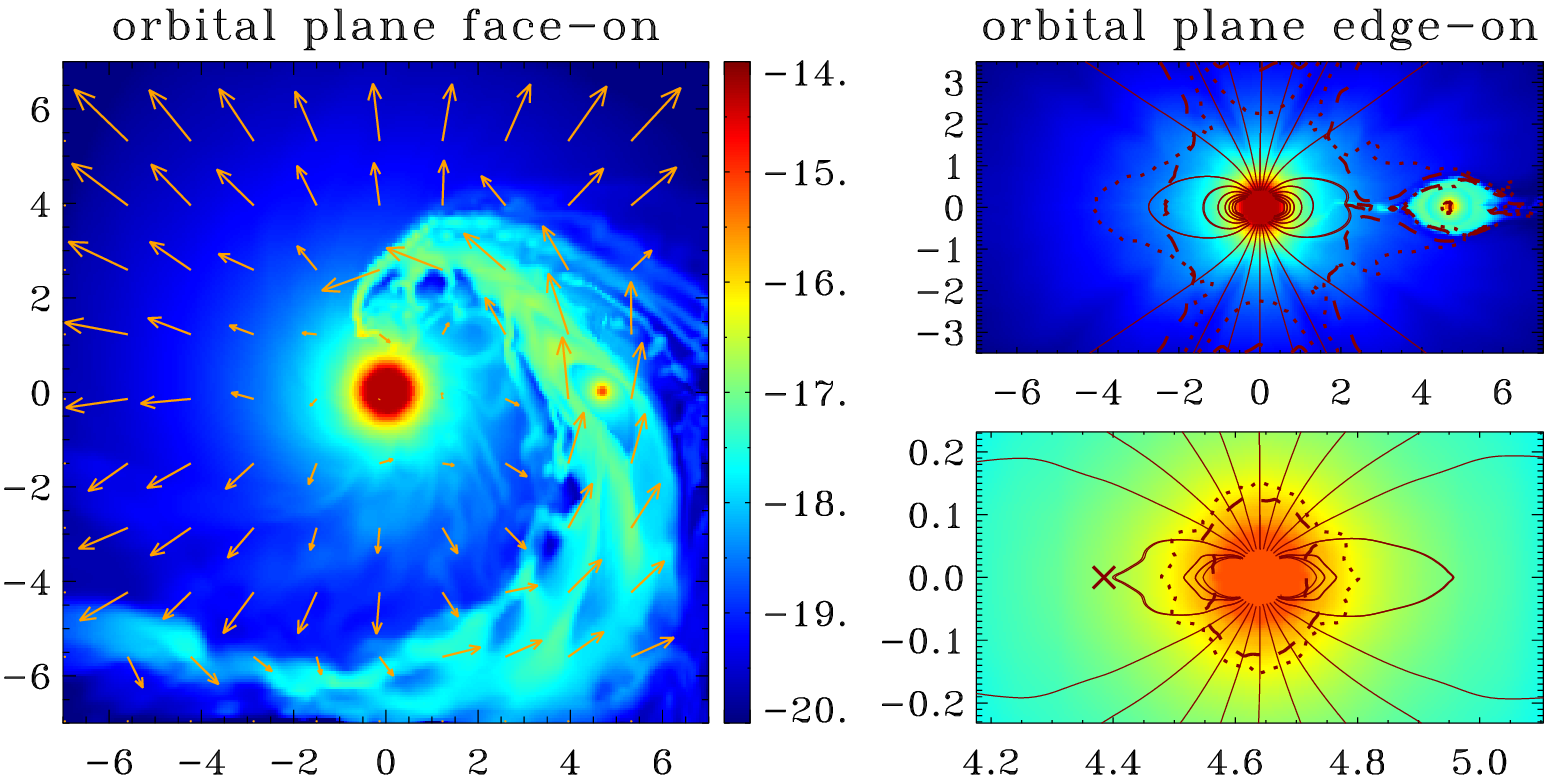}}
  \caption{
    Same as Fig.~\ref{fig:fvRb}, but for model \texttt{FvRb}.
  \label{fig:FvRb}}
\end{figure*}
\begin{figure}
  \resizebox{\hsize}{!}{\includegraphics{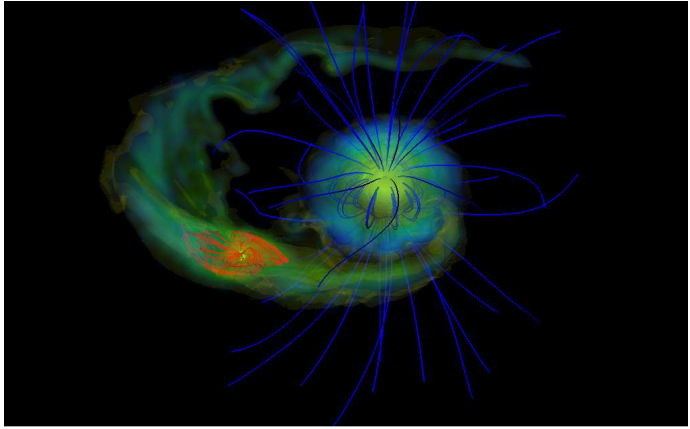}}
  \caption{
    3D representation of the density structure and of the magnetic field lines
    (blue: stellar; red: planetary) for model \texttt{FvRb}.
    The logarithmic density scale is chosen for visualization purposes and does
    not correspond to that of Fig.~\ref{fig:FvRb}.
  \label{fig:star-planet}}
\end{figure}
Figures~\ref{fig:FvRb} and~\ref{fig:star-planet} display the final
quasi-steady state of model \texttt{FvRb}.
This model differs from the reference case of Sect.~\ref{sec:tail} (model
\texttt{fvRb}, shown in Fig.~\ref{fig:fvRb}) in having a high (rather than low)
incident UV flux.
This results in the formation of a strong outflow, which opens up the
magnetosphere and becomes supersonic (bottom right panel).\footnote{The
comparatively weak magnetic field in this case implies that the  Alfv\'en
critical surface is located closer to the planet than the sonic surface.}
 
The planetary outflow propagates unperturbed over a few planetary radii, giving
rise to the eye-shaped region around the planet at $4.0\lesssim x\lesssim5.5$
(left and top right panels of Fig.~\ref{fig:FvRb}).
Beyond that region, the planetary outflow collides with the stellar wind and
forms a shock.
This happens on both the day and night sides of the Hot Jupiter on account of
the high orbital speed (which shifts the apex of the bow shock away from the
substellar point) and the axisymmetry assumption for the planetary wind.

A noteworthy feature of the interaction in this case is the accretion of part of
the shocked planetary outflow onto the host star.
The infalling material spirals for a quarter of an orbit and then impacts the
stellar surface.
The fact that this gas falls in, rather than form a torus-like structure, can be
attributed to the action of the shock and of the stellar wind, which slow down
and disrupt the flow, and to the increase in the gravitational pull of the star
as the gas spirals in.
The rest of the outflowing planetary gas is, however, pushed back and forms a
tail, as in the cases considered in the preceding subsection.
Both the inspiraling and the trailing streams are affected by the velocity shear
with the low-density stellar gas and by the outward acceleration that this
component induces, which trigger Kelvin-Helmholtz and Rayleigh-Taylor
instabilities.
These effects lead to the destruction of the tail in this case, with its
fragments being pushed outward by the ram pressure of the stellar wind (left
panel of Fig.~\ref{fig:FvRb}).
A very similar outcome is produced for model \texttt{FvRB} (not plotted), in
which, however, the stronger planetary magnetic field results in the Alfv\'en
surface lying farther away from the planet than the sonic one.

\begin{figure}
  \resizebox{\hsize}{!}{\includegraphics{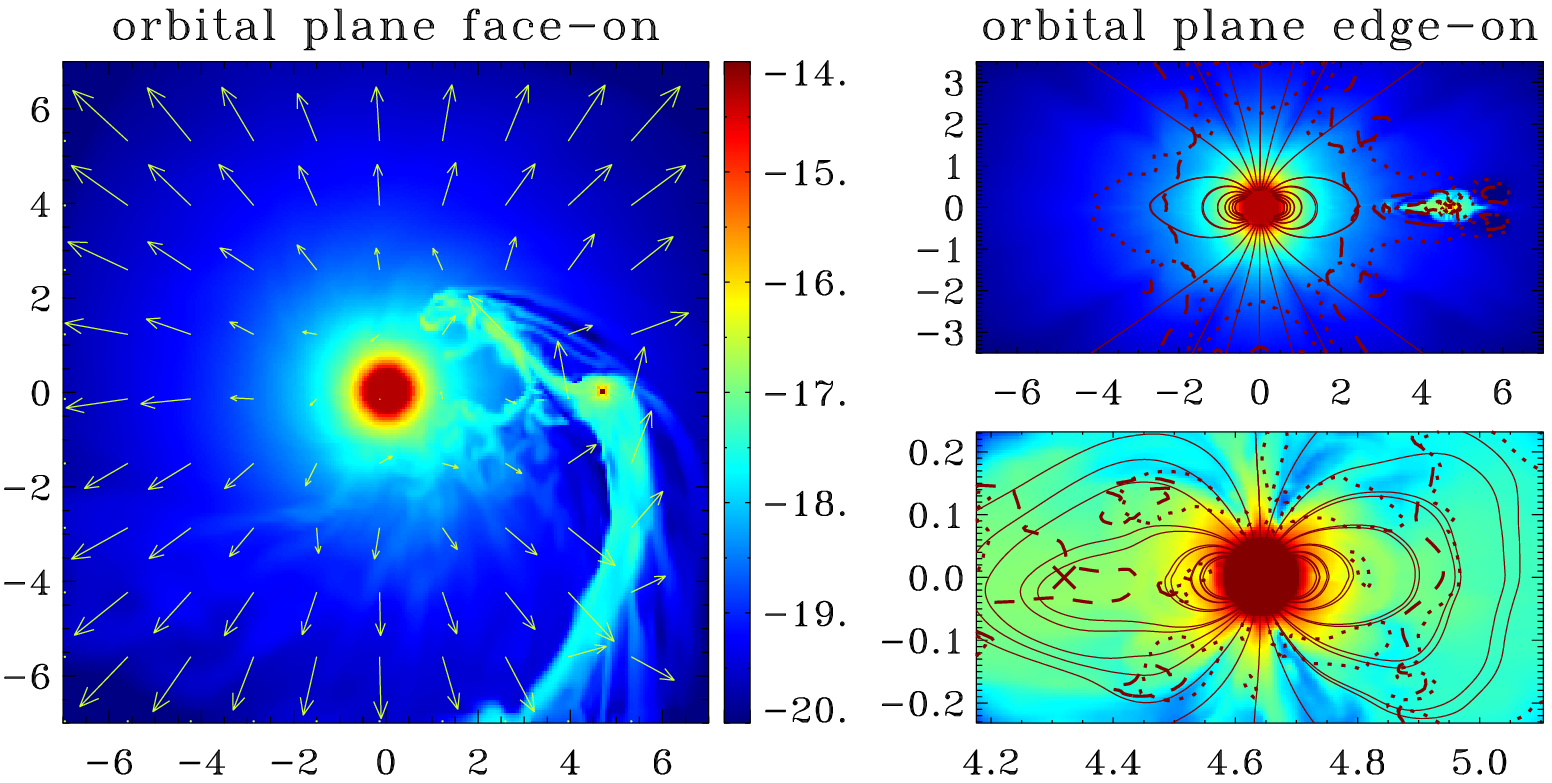}}
  \caption{
    Same as Fig.~\ref{fig:fvRb}, but for model \texttt{FVRB}.
  \label{fig:FVRB}}
\end{figure}
Accretion streams onto the stellar surface can form also in cases where the
planetary outflow remains weak and the stellar wind is stopped by the planetary
magnetic field.
This is illustrated in Fig.~\ref{fig:FVRB}, which shows the final
configuration of model \texttt{FVRB}.
This case is similar to model \texttt{FVRb}, shown in the right panel of
Fig.~\ref{fig:fvRB-FVRb}, in that a combination of high UV flux and large
escape speed generate a dense discharge from the planetary surface but only a
weak outflow.
In this case, however, the stronger magnetic field causes the interface between
the planetary magnetosphere and the shocked stellar wind to lie at a larger
distance from the planet's surface; in particular, it now lies beyond the $L_1$
point.
At that location, the gravitational pull from the star and the thermal pressure
of the dead zone combine to deform the planetary magnetic field lines in such a
way that accretion streams are formed.
These streams, however, do not flow smoothly: as is seen in the left panel of
Fig.~\ref{fig:FVRB}, their interaction with the stellar wind and the magnetic
field of the star causes their trajectories to split several times before they
reach the stellar surface.

\begin{figure}
  \resizebox{\hsize}{!}{\includegraphics{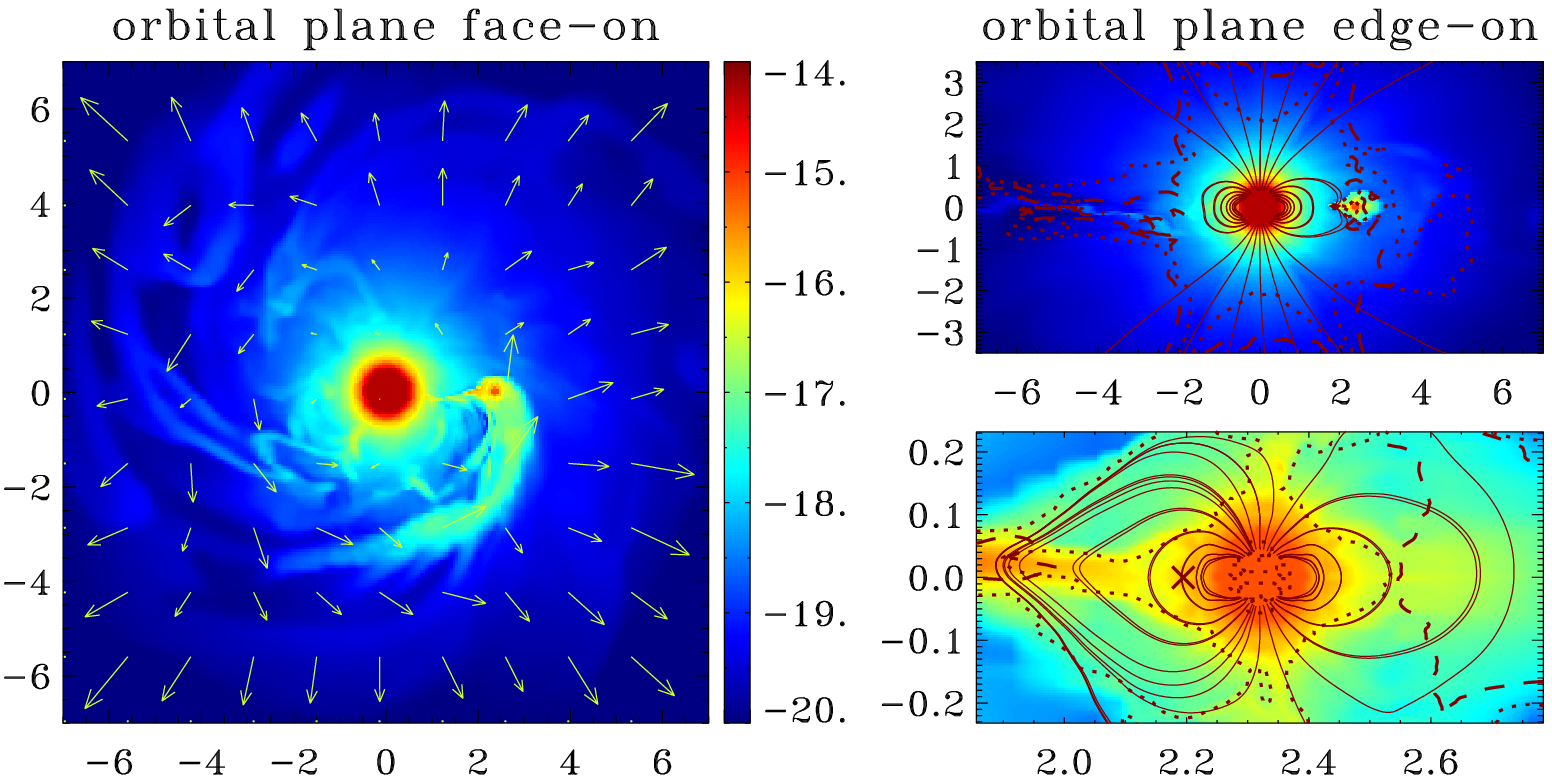}}
  \caption{
    Same as Fig.~\ref{fig:fvRb}, but for model \texttt{FvrB}.
  \label{fig:FvrB}}
\end{figure}
A similar situation characterizes our high-flux ``near'' models, as illustrated
in Fig.~\ref{fig:FvrB} with model \texttt{FvrB}.
In this case, the planetary outflow is massive enough and the $L_1$ point is
located close enough to the planet's surface that both a planetary tail and an
accretion stream are formed, resembling the behavior of model \texttt{FVRB}.
However, the different position of the planet relative to the star modifies the
evolution in this case.
First, the fragmented tail is not pushed away by the stellar outflow, which is
still weak at this location.
Instead, the stellar outflow mixes with these fragments and slows them down,
leading to their accretion by the star.
Second, the denser ambient stellar gas in this model also enhances the mixing
with the accretion stream, slowing the latter down and causing it to hit the
stellar surface close to the orbital location of the planet.
Finally, as the orbital motion of the planet is now faster than the stellar
rotation, the stream does not attain a steady-state configuration.
Instead, the magnetic field topology undergoes a continuous readjustment, with
new magnetic accretion channels forming periodically.
Model \texttt{FVrB} (not plotted) exhibits a similar morphology to that of
model \texttt{FvrB}, with accretion flows onto the star originating from both
the upstream and downstream regions of the planet, although the larger escape
speed results in less mass leaving the planet in this case.
We however find that there is no transfer of mass to the star when the UV flux
is low, irrespective of whether the escape speed is large or small
(models \texttt{fVrB} and \texttt{fvrB}, respectively; these are also not
plotted).

\section{Classification of star--planet interactions}
  \label{sec:classify}

Our simulations of the dynamical interaction between a magnetized, wind-driving
host star and a magnetized hot Jupiter that loses mass to photoevaporation were
described in Sect.~\ref{sec:results} in terms of the resulting morphological
structures.
We now attempt to distill from this phenomenology a general classification
scheme that is based on the underlying dynamical processes operating in such
systems.
We quantify the relative influence of these processes with the help of
characteristic length scales, and use the latter to identify four basic types of
interaction.
We then apply this scheme to the categorization of the models listed in
Table~\ref{tab:models}.

\subsection{Characteristic length scales}
\label{subsec:lengths}
The formation of a quasi-stationary morphological structure in the interaction
between a star and a close-in planet entails the establishment of pressure
equilibrium between the stellar and planetary plasmas at the interface
separating these two media.\footnote{This interface formally constitutes either
a contact or a tangential discontinuity.
As confirmed by our simulations, such discontinuities may be subject to
instabilities, which, among other effects, can induce mixing of the two
plasmas.}
The relevant frame of reference for considering this equilibrium is that of the
planet, and we label the pressure on the stellar side of the boundary in this
frame by $P_\mathrm{amb}$.
The characteristic distance of this surface from the center of the planet,
measured along (or close to) the line to the stellar center, is determined by
whether it is the magnetic pressure, outflow ram pressure, or thermal pressure
that dominates on the planet's side of the boundary.

If the pressure of a closed planetary magnetosphere dominates, we can estimate
the relevant scale ($R_{\rm m}$) by assuming a dipolar field,
$B _\circ(R) = B_\circ(R_\circ/R)^3$, and equating its pressure
$P_{\rm mag\, \circ} = B_\circ^2/8\pi$ to $P_\mathrm{amb}$.
This gives
\begin{equation}
  R_\mathrm{m}
    = \left(\frac{B_\circ^2R_\circ^6}{8\pi P_\mathrm{amb}}\right)^{1/6}\,.
  \label{Rm}
\end{equation}
If, however, the planetary outflow is strong enough to become supercritical by
the time it is stopped (in a shock) by colliding with the stellar gas, the
relevant scale ($R_{\rm w}$) is given, instead, by equating $P_\mathrm{amb}$ to
the ram pressure $P_{\rm ram\,\circ} = \rho_\circ v_{\mathrm{w}\circ}^2$.
Expressing the density in terms of the planetary mass loss rate
$\dot M_\circ = 4\pi R_\mathrm{w}^2\rho_{{w}\circ}v_{\mathrm{w}\circ}$, we get
\begin{equation}
  R_\mathrm{w} =    \left(\frac{\dot M_\circ v_{\mathrm{w}\circ}}
    {4\pi P_\mathrm{amb}}\right)^{1/2}\,.
  \label{Rw}
\end{equation}
Which of these pressure components (magnetic or ram) dominates the interaction
with the stellar plasma is reflected in the relative magnitudes of the
associated characteristic radii.

One could in principle also consider the contribution of the thermal pressure
$P = 2\rho k_{\rm B} T/m_{\rm p}$ of the planetary atmosphere, but in practice
its role is negligible because of its expected rapid decline with radius.
In particular, under the isothermal approximation,
$P_\circ(R') \propto \exp(R_\circ/R')$ (see Eq.~\ref{P_p}), which represents a
much faster drop than that of the magnetic pressure
($P_\mathrm{mag} \propto (R_\circ/R')^6$).
In fact, the magnetic pressure likely already dominates the thermal pressure 
at the (subsonic) base of the planetary outflow, where, using Jupiter's mass,
radius, and magnetic field ($B_\circ \simeq 4\, {\rm G}$), and the base density
($\rho_\circ \simeq 10^{-14}\,\mathrm{g\,cm}^{-3}$) and temperature
($T_\circ \simeq 10^4\,\mathrm{K}$) inferred from our high--UV-flux model (see
Fig.~\ref{fig:planet2wind}), we estimate
\begin{equation}
  \beta_\circ \equiv \frac{P_\circ}{P_{\mathrm{mag}\,\circ}}
    = 0.03\, \left( \frac{\rho_\circ}{10^{-14}\,\mathrm{g\,cm}^{-3}}\right) 
    \left(\frac{T_\circ}{10^4\, {\rm K}}\right)
    \left (\frac{B_\circ}{4\, {\rm G}}\right )^{-2}\,.
  \label{beta}
\end{equation}
Note, however, that when the stellar wind is stopped by a supercritical
planetary wind, the plasma on the planet's side of the contact discontinuity is
dominated by its thermal pressure (having passed through a shock).

The ambient pressure can be approximated by the sum of four contributions:
\begin{equation}
\label{Pamb}
  P_{\mathrm{amb}} =
    P_{\rm w*} + P_{\mathrm{mag}\,*} + P_\mathrm{ram\,*}
    + P_\mathrm{ram\,orb}\,,
\end{equation}
which represent, respectively, the stellar plasma's thermal, magnetic, and
``intrinsic'' ram pressure components, and the ram pressure associated with the
relative motion between the planet and the ambient (stellar) gas.
In this approximation, we take the direction of the stellar wind to be purely
radial and the direction of the orbital motion to be purely azimuthal in the lab
frame.
If the stellar wind is supercritical, a bow shock will form at the interaction
site.
In the limit $v_{\rm orb} \ll v_{\rm w*}$, the apex of this shock will lie along
the line connecting the centers of the star and the planet, and the shock
surface will be symmetric about this line.
However, given that the orbital speeds of hot Jupiters are typically
$\gtrsim$$100\,\mathrm{km\,s}^{-1}$ and are thus highly supersonic with respect
to the ambient gas, such a shock will form even when the stellar wind is still
subcritical when it is stopped.
In this case, however, the shock will be oriented at a finite angle to the
aforementioned line (see Eq.~\ref{eq:tail} below), and its apex will be
displaced away from this line (in the direction of the planet's
motion).\footnote{Note that the speed of the orbital motion for these systems is
of the order of the terminal speed of the stellar wind, so that, when the
interaction occurs in the supercritical region of the stellar wind, the
resulting bow shock will be significantly inclined with respect to the
instantaneous ``line of centers.''}
For typical parameters, we expect that the last two terms in Eq.~(\ref{Pamb})
constitute the dominant contributions to $P_{\rm amb}$.

One other characteristic length scale for this problem is the tidal (or Hill)
radius,
\begin{equation}
  R_\mathrm{t} = \left(\frac{M_\circ}{3M_*}\right)^{1/3}R_\mathrm{orb}\,,
  \label{R_t}
\end{equation}
which gives the distance of the $L_1$ Lagrange point from the planet's center.
This distance is relevant to the question of whether the morphology of the
planetary gas that interacts with the stellar plasma will be shaped only by the
relative motion between these two media (through the effect of the ram pressure
terms in Eq.~\ref{Pamb}) or whether the stellar gravitational field will also
play a role.
It can be expected that at least some of the planet's plasma will be accreted
onto the stellar surface if $R_{\rm t} < \max{(R_{\rm m},R_{\rm w})}$, but that
essentially all of the planet's gas will remain confined to the vicinity of its
orbital radius if this inequality is not satisfied.

Before concluding this subsection, we note that the components of $P_{\rm amb}$
can be expressed in terms of the stellar parameters just as was done above for
the planet's pressure components.
In particular, using $\dot M_* = 4\pi R^2\rho_{\mathrm{w}*}v_{\mathrm{w}*}$,
$B_*(R) = B_*(R_*/R_{\rm orb})^3$, and the isothermal-gas assumption, we can
express the terms appearing on the right-hand side of Eq.~(\ref{Pamb}) in the
form
\begin{eqnarray}
  P_{\rm w*} &=&
    \frac{2\rho_{\mathrm{w}*}k_\mathrm{B}T_{\mathrm{w}*}}{m_\mathrm{p}}
    = \frac{\dot M_*k_\mathrm{B}T_{\mathrm{w}*}}
    {2\pi m_\mathrm{p}R_\mathrm{orb}^2v_\mathrm{w*}}\,, \\
  P_\mathrm{mag\,*} &=&
    \frac{B_*^2}{8\pi}\left(\frac{R_*}{R_\mathrm{orb}}\right)^6\,, \\
  P_\mathrm{ram\,w*} &=& \rho_{\mathrm{w}*}v_\mathrm{w*}^2
    = \frac{\dot M_*v_\mathrm{w*}}{4\pi R_\mathrm{orb}^2}\,, \\
  P_\mathrm{ram\,orb} &=& \rho_{\mathrm{w}*}v_\mathrm{orb}^2
    = \frac{\dot M_*v_\mathrm{orb}^2}{4\pi R_\mathrm{orb}^2v_\mathrm{w*}}\,.
\end{eqnarray}
Thus, the values of the three characteristic length scales ($R_\mathrm{m}$,
$R_\mathrm{w}$, and $R_\mathrm{t}$) can be estimated from a given set of
planetary ($R_\circ$, $M_\circ$, $\dot M_\circ$, $v_{\mathrm{w}\circ}$,
$B_\circ$), orbital ($R_\mathrm{orb}$, $v_\mathrm{orb}$), and
stellar ($R_*$, $M_*$, $\dot M_*$, $v_{\mathrm{w}*}$,
$T_{\mathrm{w}*}$, $B_*$) parameters.

\subsection{Types of interaction}
\label{subsec:types}

\begin{figure*}
  \begin{center}
  \resizebox{\hsize}{!}{\includegraphics{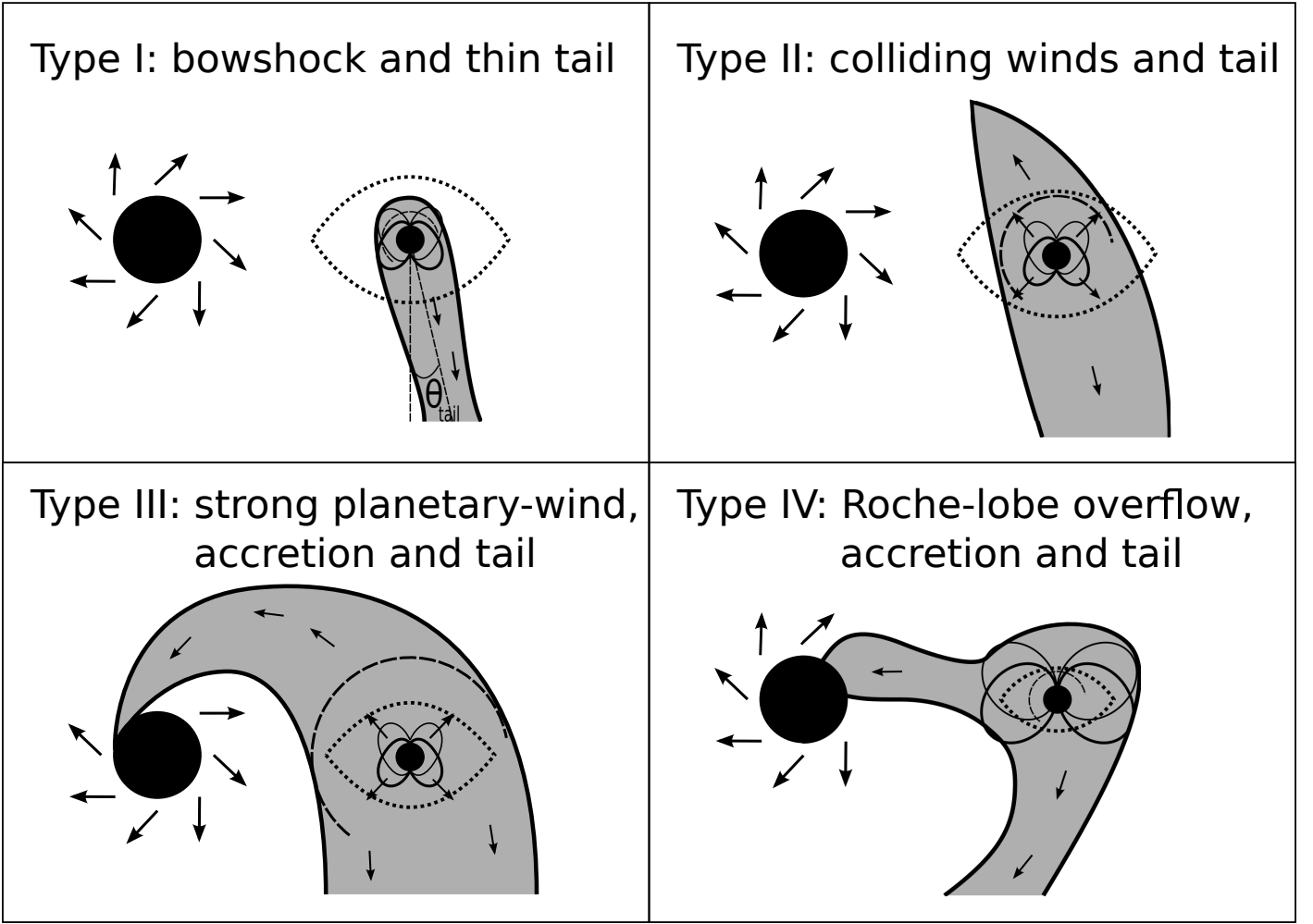}}
  \caption{
    Schematic of the different possible star--planet interactions,
    showing face-on views in the orbital plane of the four distinct
    morphological structures (denoted by the Latin numerals I, II, III, and~IV)
    as they appear in the planet's frame.
    The large and small solid disks (on the left and right sides of each panel)
    represent the star and the Hot Jupiter, respectively, the shaded areas
    highlight material that flowed out of the planet, the arrows indicate gas
    motions, and the closed loops sketch the planetary magnetosphere.
    The solid circular arcs in panels~I and~IV have a radius $R_{\rm m}$, the
    distance where the magnetospheric pressure equals the total ambient
    pressure, whereas the dashed circular arcs in panels~II and~III mark the
    distance $R_\mathrm{w}$ where the ambient pressure equals the ram pressure
    of the planetary wind.
    The dotted line indicates the contour of the effective (gravity plus
    centrifugal) potential that crosses the Lagrange point $L_1$ (at a distance
    $R_\mathrm{t}$ from the planet). 
    The proposed classification scheme is based on the relative ordering of
    $R_\mathrm{m}$, $R_\mathrm{w}$, and $R_\mathrm{t}$.
    \label{fig:sketch}}
  \end{center}
\end{figure*}

We base our classification of the flow structures that arise from the
hydromagnetic interaction between a close-in giant planet and its host star on
the relative magnitudes of the three characteristic length scales considered in
Sect.~\ref{subsec:lengths}: $R_\mathrm{m}$, $R_\mathrm{w}$, and
$R_\mathrm{t}$.
The four basic types of interaction that we identify in this way are sketched in
Fig.~\ref{fig:sketch} and discussed below.
We note, however, that any given system may not fall exclusively under a single
category.
This could be due to the planet having an eccentric orbit, to the stellar wind
being nonaxisymmetric, or to the presence of multipolar components in the
stellar magnetic field, which could each lead to variations in the values of
$R_{\rm m}$ and $R_{\rm w}$ on a timescale of $\sim$days.
A stellar magnetic cycle would be likely to induce variations on timescales of
$\sim$years in the properties of the stellar magnetic field, the stellar wind,
and the stellar UV flux.
Changes on even longer timescales could result from the effects of stellar
evolution.

\medskip\noindent
\underline{Type I: $R_\mathrm{t} > R_\mathrm{m} > R_\mathrm{w}$}

\medskip\noindent
Type I interactions occur when the planetary outflow is weak (corresponding to
either a low irradiating flux or a large escape speed), so that the stellar
plasma is intercepted by the planetary magnetic field (i.e.,
$R_{\rm m} > R_{\rm w}$).
As we noted in Sect.~\ref{subsec:lengths}, the relative motion between the
planet and the ambient gas is typically high enough to lead to the formation of
a bow shock upstream of the planet.
The shocked stellar flow sweeps back the planetary magnetic field and diverts
the planetary outflow in the downstream direction, leading to the formation of a
thin planetary tail.
The stronger the planetary field, the larger its pressure --- hence the farther
away from the planetary surface does the shock appear ($R_{\rm m}$ is larger),
and the wider the tail that is produced.
The tail gas remains confined to the vicinity of the planet's orbit because
$R_{\rm t} > R_{\rm m}$.

The orientation of the tail is determined by the direction of the incident
stellar wind as seen in the corotating frame at the apex of the bow shock.
Approximating the wind velocity to be nearly radial in the lab frame and of
magnitude $v_\mathrm{w}$ (see Fig.~\ref{fig:initial}), the angle that the tail
makes with respect to the tangent to the planet's orbit (indicated on the top
left drawing in Fig.~\ref{fig:sketch}) can be approximated by
\begin{equation}
\label{eq:tail}
  \theta_\mathrm{tail} \approx \arctan{\frac{v_\mathrm{w\,*}}{v_\mathrm{orb}}} =
    \arctan{\left(\frac{v_\mathrm{w\,*}^2R_\mathrm{orb}}{GM_*}\right)^{1/2}}\,.
\end{equation}
\citep[see also][]{Vidotto+10}. 
In the limit of an extended and static stellar corona ($v_\mathrm{w*} = 0$), the
tail will trail the trajectory, forming a torus around the star.
In the other limit, when a strong wind hits a comparatively distant planet, the
tail will be almost perpendicular to the orbit, and so will lie nearly along the
line of sight during transits.
Note that we did not include the effect of radiation pressure, which is the
dominant force in the formation of comet-type tails in close-in rocky planets
\cite[e.g.,][]{Rappaport+12, Rappaport+14}.
In the case of the gaseous giant planets considered in this paper, radiation
pressure on Ly$\alpha$ lines could potentially play a role, but its contribution
would be negligible if (as we suggest below) the gas in the tail is highly
ionized.
Furthermore, even in regions where the gas may still be only partially ionized
(for example, near the base of the planetary outflow), the amount of gas
affected by this process would be strongly limited by opacity effects
\citep[e.g.,][]{TremblinChiang13}.
It is, however, conceivable that radiation pressure on resonance lines such as
those of H~I and Mg~I could contribute to the blueshifted spectral features
observed in the measured absorption profiles of these lines \citep[e.g.,][and
references therein]{BourrierLecavelierDesEtangs13,Bourrier+14}.

Our simulations indicate that the gas that dominates the column density between
the Hot Jupiter and the bow shock is of planetary origin, rather than the
shocked stellar plasma.
We obtain a representative value for the hydrogen column density of this
component of $N_{\rm H\,\circ} \approx 8\times 10^{15}\, {\rm cm}^{-2}$ (using a
typical hydrogen number density of
$n_{\rm H\,\circ} \approx 6\times 10^5\, {\rm cm}^{-3}$ and a path length of
$\sim 2\, R_\circ$, based on the simulation results for model \texttt{fVRb}; see
Fig.~\ref{fig:fVRb}).
This column can be compared with the equilibrium column of an ionized slab with
this density,
\begin{eqnarray}
  \label{eq:N_II}
  N_{\rm H II} &\approx&
   \frac{F_\mathrm{UV}}{\alpha_\mathrm{B}n_{\rm H\,\circ}E_\mathrm{UV}} \\
   &=& 9\times 10^{19}
   \left(\frac{F_{\rm UV}}{\rm 450\,erg\,cm^{-2}\,s^{-1}}\right)
   \left(\frac{n_{\rm H\, \circ}}{6\times 10^5\,{\rm cm}^{-3}}\right)^{-1}\,
   {\rm cm}^{-2},\nonumber
\end{eqnarray}
where
$\alpha_\mathrm{B} \simeq 2.6 \times 10^{-13}\,\mathrm{cm}^3\,\mathrm{s}^{-1}$
is the Case B recombination coefficient, and where (following
\citealt{MurrayClay+09}) we adopted a characteristic ionizing photon energy of
$E_{\rm UV} = 20\,$eV.
This estimate indicates that, even for a comparatively low ionizing flux, the
entire planetary gas in the interaction region would be fully ionized, implying
that this component would not be detectable in Ly$\alpha$.
A possible caveat to this conclusion may arise when the configuration is not in
quasi-static equilibrium, and in particular if the ionization time, 
\begin{eqnarray}
  \label{eq:t_ion}
  t_{\rm ionize} &\approx&
    \left(\frac{E_{\rm UV}}{\sigma_0 F_{\rm UV}}\right)
    \left(\frac{E_{\rm UV}}{13.6\,{\rm eV}}\right)^3\\
    &=& 3.8\times 10^4 \left(\frac{E}{20\,{\rm eV}}\right)^4
    \left(\frac{F_{\rm UV}}{{\rm 450\,erg\,cm^{-2}\,s^{-1}}}\right)^{-1}
    \,{\rm s}\nonumber
\end{eqnarray}
(where $\sigma_0 \simeq 6\times 10^{-18}\, {\rm cm}^2$ is the hydrogen
photoionization cross section at the Lyman edge), is longer than the travel time
of the planetary gas to the interaction region.
However, this is probably not relevant to Type~I configurations on account of
the low velocity of the outflowing gas.
The planetary gas in Type~I systems might nevertheless be detectable in other
absorption lines, such as the UV h and k resonance lines of Mg~II.

\medskip\noindent
\underline{Type II: $R_\mathrm{t} > R_\mathrm{w} > R_\mathrm{m}$}

\medskip\noindent
Type II interactions occur when the planetary outflow is comparatively strong
(corresponding to either a high irradiating flux or a small escape speed), so
that the stellar plasma is intercepted by the planetary outflow (i.e.,
$R_{\rm w} > R_{\rm m}$).
We again expect this plasma (which moves with a velocity of a few hundred
km\,s$^{-1}$ with respect to the planet) to pass through a bow shock, and, after
being shocked, to sweep back the planetary outflow into a tail.
Because of the larger momentum flux in the planetary outflow in comparison with
the Type~I case, Type~II tails are wider than their Type~I counterparts for a
given value of $B_\circ$. However, just as in the case of Type~I interactions,
the tail gas remains confined to the vicinity of the planet's orbit, in this
case because $R_{\rm t} > R_{\rm w}$.
Its tilt angle ($\theta_{\rm tail}$) with respect to the orbit can be similarly
estimated using Eq.~(\ref{eq:tail}).

Since in this case the planetary outflow is supercritical by the time it
collides with the stellar plasma on the day side of the planet (at which point
its speed $v_{\rm\,w \circ}$ is $\lesssim 30\,{\rm km\,s}^{-1}$, on the order of
$v_\mathrm{esc\,\circ}$), it is also decelerated in a shock.
The two shocks are separated by a contact discontinuity, across which the total
(thermal + magnetic) pressure is the same.
This, in turn, implies that the normal components (with respect to the contact
discontinuity surface) of the ram pressures of the two flows are approximately
equal in the frame of the contact discontinuity (which, in a steady state, is
the same as that of the planet).
As a rough estimate, we write
\begin{equation}
  \rho_{\mathrm{w}\,*} (v_{\mathrm{w}\,*}^2 + v_\mathrm{orb}^2)
  \sim \rho_{\mathrm{w}\,\circ} v_{\mathrm{w}\,\circ}^2\,,
  \label{eq:ram}
\end{equation}
where, as we already noted, $v_{\mathrm{w}\,*}$ is typically of the order of
$v_\mathrm{orb}$ for a Hot Jupiter.
The estimate (\ref{eq:ram}) implies that the ratio of the preshock densities in
the interaction region is
$\rho_{\mathrm{w}\,*}/\rho_{\mathrm{w}\,\circ}
\sim v_{\mathrm{w}\,\circ}^2/(v_{\mathrm{w}\,*}^2 + v_\mathrm{orb}^2)$, which
for $v_{\mathrm{w}\,\circ} \sim 0.1\, v_{\mathrm{w}\,*}$ is $\sim$$0.01$.
The ratio of the corresponding postshock densities need not be the same on
account of differences in the shock strength and the postshock cooling
efficiency between the two winds, but it would likely still be $\ll 1$.
It also follows from Eq.~(\ref{eq:ram}) that the ratio of the mass inflow rates
into the interaction zone,
$\sim \rho_{\mathrm{w}\,*}(v_{\mathrm{w}\,*}^2 + v_\mathrm{orb}^2)^{1/2}
/(\rho_{\mathrm{w}\,\circ} v_{\mathrm{w}\,\circ})$, is
$\sim v_{\mathrm{w}\,\circ}/(v_{\mathrm{w}\,*}^2 + v_\mathrm{orb}^2)^{1/2}$
($\sim$$0.1$ for our adopted values).
This suggests that, as in the Type~I case, the postshock density and column
density in the interaction region is dominated by the planetary material.
Although our simulations have not produced examples of a Type~II interaction, we
can use the simulation results for model \texttt{FvRb} (Fig.~\ref{fig:FvRb}),
which represents an example of a Type~III interaction that involves a
comparatively massive planetary outflow (see Table~\ref{tab:models} and
Sect.~\ref{subsec:classify}), to infer an upper bound of
$\sim$$10^{-16}\,\mathrm{g\,cm}^{-3}$ on the typical postshock planetary density
in this case.
We note in this connection that a supercritical planetary outflow is also
shocked on the night side of the planet when it hits the gas that had been
shocked further upstream and subsequently dragged backward around the
planet.\footnote{The detailed morphology of the night-side region may change if
the assumption of equivalent outflows from the two sides of the planet is
modified.
However, we do not expect our basic classification scheme to be sensitive to
these details.} 
Moving farther out from the star in that direction, one again encounters
unshocked stellar-wind gas that expands nearly radially, so the density there
rapidly becomes very low.

Our estimate of the density of the planetary gas in the interaction region is
two orders of magnitude larger in this case than the typical value we adopted
for the Type~I interaction ($\sim$$10^{-16}\,\mathrm{g\,cm}^{-3}$, associated
with a high flux value, vs. $\sim$$10^{-18}\,\mathrm{g\,cm}^{-3}$, which was
obtained for a low UV flux).
Using Eq.~(\ref{eq:N_II}), we deduce a nominal ionized column of
$N_{\rm H II} \approx 10^{21}\, {\rm cm}^{-2}$ for these parameters, as compared
with a total column of $N_{\rm H\, \circ} \lesssim 8 \times 10^{17}$, indicating
that, in this case too, the interacting gas can be expected to be fully ionized
if in equilibrium.
The equilibrium assumption could be questioned since the gas is clearly in
motion for Type~II configurations; however, for the high flux value that
characterizes the flow in model \texttt{FvRb}, the nominal ionization time
inferred from Eq.~(\ref{eq:t_ion}) (a few tens of seconds) is much shorter than
any relevant flow time.
In addition, the shock heating of the planetary outflow can also contribute to
the ionization of this gas.\footnote{The postshock temperature scales as the
square of the upstream gas velocity normal to the shock front; one can therefore
expect temperatures of a few times $10^4\,$K for the shocked planetary flow, as
compared with a few million degrees for the shocked stellar plasma.}
It is therefore likely that much of the gas that participates in a Type~II
interaction is ionized, although this conclusion needs to be checked explicitly
for each given set of parameters.

\medskip\noindent
\underline{Type III: $R_\mathrm{w} > R_\mathrm{m}$ \&
$R_\mathrm{w} > R_\mathrm{t}$}

\medskip\noindent
Type~III interactions are similar to those of Type~II in that they involve a
strong planetary wind ($R_{\rm w}>R_{\rm m}$).
However, in this case the planetary outflow, after being stopped in a shock, is
only partly swept back into a tail, with another part responding to the
gravitational pull of the star (which is felt because $R_{\rm w}>R_{\rm t}$) and
penetrating the stellar plasma ahead of the planet, forming a stream that
spirals in and accretes onto the star.\footnote{If the planet orbits close
enough to the critical surface of the stellar wind (as defined in the corotating
frame), it may be possible for the stellar outflow to encounter the boundary of
the stream while it is still subcritical, in which case no bow shock will form.
However, a shock may still form when the stellar wind intercepts the planetary
tail if it is already supercritical at that location (see, e.g.,
Fig.~\ref{fig:FvRb}).}

The accretion stream exhibits a complex, fragmented morphology that arises from
Kelvin-Helmholtz and Rayleigh-Taylor instabilities --- triggered, respectively,
by the tangential velocity shear and by the radial acceleration exerted by the
stellar wind gas on the denser planetary material.
The stellar gravity further contributes to the development of the latter
instability on account of the ``heavy on top of light'' density stratification.
These processes, and possibly also the magnetic stresses acting on the inflowing
plasma, fragment the inspiraling stream, resulting in the formation of multiple
accretion filaments that hit the stellar surface at different spots.
The general location where the accretion stream impacts the stellar surface is
determined by the stellar and planetary wind parameters and by the orbital
distance of the planet, but it is typically well ahead of the instantaneous
sub-planetary point (the phase difference being $\sim$$90^\circ$ in the example
shown in Fig.~\ref{fig:FvRb}). 
We do not observe the formation of a circumstellar disk in our simulations,
evidently because the drag exerted by the stellar outflow and the mixing between
the two plasmas result in efficient removal of angular momentum from the
inspiraling gas.

There have been several claims in the literature for enhanced activity on the
host star --- manifested in optical chromospheric emission lines (particularly
Ca~II), transition-region FUV emission lines (such as Si~IV), and coronal X-ray
emission --- that might be attributable to an interaction with a close-in planet
\citep[e.g.,][]{Shkolnik+03, Shkolnik+04, Shkolnik+05, Shkolnik+08, Walker+08,
Pillitteri+10, Pillitteri+11, Pillitteri+14, Pillitteri+15}.
The accretion streams produced in Type~III interactions could potentially
trigger such events.
In particular, they may be associated with active regions that are inferred to
lie significantly ahead of the sub-planetary point --- as in HD~179949 and
$\tau$~Boo, where a phase difference of $\sim$$70^\circ$ was inferred from
measurements of the Ca~II K line \citep{Shkolnik+08, Walker+08}, and in
HD~189733, where a lead phase of $\sim$$70^\circ$--$90^\circ$ was deduced from
X-ray and FUV observations \citep{Pillitteri+14, Pillitteri+15}.
An alternative interpretation of these phenomena that has been discussed in the
literature involves a direct magnetic interaction of the type observed in
planetary moons (such as Io) in the solar system.
Note, however, that in their simplest form, such interactions are predicted to
occur near the sub-planetary point.
Furthermore, the energy flux generated in such an interaction in a source like
HD~179499 is calculated \citep{Saur+13} to be over two orders of magnitude lower
than the observationally inferred value \citep{Shkolnik+05}.

\medskip\noindent
\underline{Type IV: $R_\mathrm{m} > R_\mathrm{w}$ \&
$R_\mathrm{m} > R_\mathrm{t}$}

\medskip\noindent
This case is analogous to the Type~III one in that the interaction region lies
outside the planet's Hill radius, resulting in the accretion of planetary gas
onto the stellar surface.
However, just as a Type~III interaction could be regarded as a special case of a
Type~II one, with both involving a strong planetary outflow, a Type~IV
interaction also bears similarity to a Type-I interaction in having a
comparatively weak planetary wind ($R_{\rm m}>R_{\rm w}$).
In this case the planetary magnetosphere, which is loaded by the subsonic gas
evaporated by the stellar radiation field, intercepts the stellar plasma beyond
the $L_1$ point, where the stellar gravity is the dominant force acting on the
planetary gas.
This situation resembles the classic Roche-lobe overflow picture, although both
the planetary magnetic field and the outflowing, magnetized  stellar plasma also
play a role in shaping the accretion flow morphology.
In particular, when the orbital radius is small enough, the relative motion
between the stellar gas and the planet may be supercritical and therefore lead
to the formation of a bow shock ahead of the planet, as in the Type~I case, even
as mass transfer from the planet to the star takes place in the vicinity of the
substellar region.
However, no bow shock forms on the star-facing side in our ``near'' models
because the stellar magnetic field (and hence the Alfv\'en speed) at $R_{\rm m}$
is large enough in these cases to keep the relative motion between the stellar
gas and the planet subcritical (see Fig.~\ref{fig:FvrB}).

The structure of the accretion flow and its dynamical implications are similar
in this case to the situation in Type~III interactions.
However, because of the absence of a strong planetary outflow in the Type~IV
case, the accretion stream does not exhibit the spiral structure observed in
Type~III interactions and instead maintains a nearly linear shape, hitting the
stellar surface near the sub-planetary point.
This type of interaction may thus be relevant to the interpretation of enhanced
stellar chromospheric/X-ray activity that is inferred to occur near the
sub-planetary point (and could provide an alternative to the ``direct magnetic
interaction'' scenario also in this case).

A Roche-lobe overflow model was previously proposed for the interpretation of
early ingress indications in UV absorption-line observations of the short-period
($R_{\rm orb} = 0.023\,$AU) Hot Jupiter WASP-12b \citep{Fossati+10}.
In particular, \citet{Lai+10}, considered the possible contributions of an
accretion stream and of an inner accretion disk, and also speculated on
absorption in the interaction region between a stellar wind and the planetary
magnetosphere.
They did not, however, include the effect of the stellar radiation heating of
the planet and the influence of the stellar plasma on the accretion stream.
\citet{Bisikalo+13} carried out a hydrodynamical simulation and drew attention
to the interaction between the accretion stream and the stellar outflow, but
they did not consider the effect of a stellar and/or a planetary magnetic field.
The results presented in this paper provide a more general framework for
modeling systems like WASP-12, and may also aid in modeling other types of
close-in planets where Roche-lobe overflow could play a role \citep[e.g.,][]
{Valsecchi+14}.

\subsection{Classification of the simulations}
\label{subsec:classify}

\begin{table}
  \caption{
    Classification of the models listed in Table~\ref{tab:models}.
  \label{tab:types}}
  \centering 
  \begin{tabular}{lc}
    \hline
    \hline
    Model & Interaction type \\
    \hline
    \texttt{FvRB} & III \\
    \texttt{FvRb} & III \\
    \texttt{FvrB} & IV  \\
    \hline
    \texttt{fvRB} & I   \\
    \texttt{fvRb} & I   \\
    \texttt{fvrB} & I   \\
    \hline
    \texttt{FVRB} & IV  \\
    \texttt{FVRb} & I   \\
    \texttt{FVrB} & IV  \\
    \hline
    \texttt{fVRB} & I   \\
    \texttt{fVRb} & I   \\
    \texttt{fVrB} & I   \\
    \hline
  \end{tabular}
\end{table}

The application of our proposed classification scheme to the models that we have
simulated is presented in Table~\ref{tab:types}.
Note that Type~II interactions are not represented in this table.
This interaction is intermediate between the Type~I and Type~III cases --- for
which we do have examples --- in that the planetary outflow must be strong
enough to cause $R_{\rm w}$ to exceed $R_{\rm m}$ but not so strong that it will
also exceed $R_{\rm t}$.
Computational constraints prevented us from exploring a wider parameter space
that would have encompassed this interaction.

\section{Summary \& conclusions}
  \label{sec:summary}

In this paper we study the different types of star-planet interactions that can
occur in systems harboring Hot Jupiters.
We perform a series of parametrized 3D MHD numerical simulations, incorporating
the star, the planet, and consistent stellar and planetary winds.
Based on the results of a grid of models, we propose a classification of the
interactions into four general categories (which we label I--IV; see
Fig.~\ref{fig:sketch}).
We describe in detail the dynamical features exhibited in each case, and provide
expressions for the three characteristic length scales that underlie our
classification scheme in terms of the system's physical parameters.
We also briefly discuss general characteristics of these interactions that could
have a bearing on their observational properties. 

The main conclusions of our analysis can be summarized as follows:
\begin{enumerate}
\item
Except very close to the star, where the stellar magnetic field (and the
associated Alfv\'en speed) might be large enough to render the flow subcritical,
the stellar outflow intercepted by the planet is stopped in a bow shock.
The shock forms even when the stellar wind is not yet fully accelerated because
of the high ($\gtrsim 100\, {\rm km\,s}^{-1}$) azimuthal velocity component of
the stellar gas in the planet's frame.
\item
In Type~I interactions, the stellar plasma is intercepted by the planetary
magnetosphere, which, in turn, is swept back into a plasma tail that is blown
away by the stellar wind.
\item
In interactions of Type~II and~III, the stellar plasma is intercepted by a
strong planetary outflow that opens up the magnetosphere.
The shocked planetary plasma is dragged backward by the stellar outflow, forming
a wide tail that is fragmented by dynamical instabilities.
In the Type~II case, the shocked planetary plasma lies inside the Roche lobe and
remains confined to the vicinity of the orbital radius.
In contrast, in the Type~III case the interaction occurs outside the Roche
surface, resulting in part of the shocked planetary plasma forming an accretion
stream that spirals toward the star until it impacts the stellar surface.
The dynamical interaction of this stream with the magnetized stellar wind
enables this gas to lose energy and angular momentum and fall in, instead of
forming a plasma torus.
The stream is also subject to dynamical instabilities, which cause it to split
into multiple accretion filaments that hit the stellar surface at different
spots.
\item
In Type~IV interactions the stellar outflow is intercepted by the planetary
magnetic field outside the Roche radius.
This situation resembles a classic Roche-lobe overflow, although it differs in
detail on account of the presence of the planetary magnetic field and of the
magnetized stellar outflow. In contrast with the Type~III accretion stream,
which reaches the stellar surface well ahead of the sub-planetary point, in this
case the planetary gas falls in nearly radially.
\item
In general, the density and column density in the interaction region are
dominated by the planetary gas.
Typically, this gas can be expected to be fully ionized, implying that detection
through absorption features of neutral species such as H~I and Mg~I is unlikely. 
\end{enumerate}

\begin{acknowledgements}
We thank Sean Matt for his helpful guidance in the implementation of the stellar
wind models.
TM is grateful for the hospitality and support of the Harvard-Smithsonian Center
for Astrophysics, and for the very useful input from I.~Pillitteri and S.~Wolk.
We also acknowledge the constructive comments of the anonymous referee.
This work was completed with resources provided by the University of Chicago
Research Computing Center and was supported in part by NASA ATP grant 
NNX13AH56G and NSF grant AST-0908184.
\end{acknowledgements}

\bibliographystyle{aa}
\bibliography{paper}

\end{document}